\documentclass[10pt,journal]{IEEEtran}
\usepackage{lscape}
\usepackage{rotating}

\usepackage[utf8]{inputenc}
\usepackage[english]{babel}
\usepackage[babel=true]{csquotes}

\usepackage{amsmath, amssymb, amsthm, wasysym, mathtools, esvect, dsfont} % math packages
\usepackage{url, hyperref, cleveref} % ref packages
\usepackage{tabularx, multirow, booktabs, colortbl} % table packages

\usepackage{graphicx}
\usepackage{xcolor}
\usepackage{tikz}
\usetikzlibrary{automata, arrows, shadows, decorations.text, circuits.logic.US, matrix, shapes.geometric, patterns} 

\usepackage{listings}
\usepackage{algorithm}
\usepackage[noend]{algpseudocode}
\algrenewcommand\algorithmicrequire{\textbf{\ \ Input:}}
\algrenewcommand\algorithmicensure{\textbf{Output:}}

\usepackage[nolist]{acronym}
\usepackage{float}
\usepackage{xspace}
\usepackage{comment}
\usepackage{siunitx}
\usepackage{adjustbox}
\usepackage{subcaption}

\makeatletter
\newcommand{\removeifnextchar}[2]{%
    \begingroup
    \ltx@LocToksA{\endgroup#2}%
    \ltx@ifnextchar@nospace{#1}{%
        \def\next{\the\ltx@LocToksA}%
        \afterassignment\next
        \let\scratch= %
    }{%
        \the\ltx@LocToksA
    }%
}
\newcommand{\etal}{\protect\removeifnextchar{.}{et~al.\@ifnextchar.{}{~}}}
\makeatother

\ifdefined\algorithmautorefname
\else
\newcommand{\algorithmautorefname}{Algorithm}
\fi 

\ifdefined\definitionautorefname
\else 
\newcommand{\definitionautorefname}{Definition}
\fi

\addto\extrasenglish{  
    \ifdefined\algorithmautorefname
    \renewcommand{\algorithmautorefname}{Algorithm}
    \fi 
    \ifdefined\definitionautorefname
    \renewcommand{\definitionautorefname}{Definition}
    \fi

}

\newcommand{\Figure}[1]{\autoref{#1}}

\newcommand{\Section}[1]{\autoref{#1}}
\newcommand{\Table}[1]{\autoref{#1}}

\newcommand{\revise}[1]{\textcolor{blue}{#1}}

\newcommand{\CPP}[1][]{$\text{C\hspace{-.25ex}}^{_{_{_{++}}}}\ifthenelse{\equal{#1}{}}{}{\text{\hspace{-.625ex}#1}}$\xspace}

\newcommand{\circled}[2][]{
    \tikz[baseline=(char.base)]{
        \node[shape=circle,draw,inner sep=1pt,fill=black]
        (char) {\phantom{\ifblank{#1}{#2}{#1}}};
        \node[text=white] at (char.center) {\makebox[0pt][c]{\textbf#2}};}\xspace}
\robustify{\circled}

\newcommand{\emu}{\textsf{M-ulator}\xspace}
\newcommand{\fsim}{\textsf{ARMORY}\xspace}
\newcommand{\model}[1]{\textsf{#1}\xspace}
\newcommand{\arm}[1]{\mbox{{ARMv}#1{-M}}}

\renewcommand{\revise}[1]{#1}

\def\checkmark{\tikz\fill[scale=0.4](0,.35) -- (.25,0) -- (1,.7) -- (.25,.15) -- cycle;} 
\def\crossout{\tikz\fill[scale=0.3](0,0) -- (.4,.5) -- (0,1) -- (.5,.6) -- (1,1) -- (.6,.5) -- (1,0)  -- (.5,.4) -- cycle ;} 

\begin{acronym}
    \acro{DFA}{Differential Fault Analysis}
    \acro{IC}{Integrated Circuit}
    \acro{IoT}{Internet of Things}
\end{acronym}

\title{\textsf{ARMORY}: Fully Automated and Exhaustive Fault Simulation on ARM-M Binaries}

\author{Max Hoffmann, Falk Schellenberg, Christof Paar \IEEEmembership{Fellow, IEEE}%
	\thanks{M. Hoffmann was with the Horst G\"ortz Institute for IT-Security, Ruhr University Bochum, Germany.}
	\thanks{M. Hoffmann, F. Schellenberg, and C. Paar were with the Max Planck Institute for Security and Privacy, Germany.}}

\begin{document}

\maketitle

%\keywords{Fault Simulation \and Fault Analysis \and ARM}

%-------------------------------------------------------------------------------

% !TeX root = main.tex

\begin{abstract}
    Embedded systems are ubiquitous.
    However, physical access of users and likewise attackers makes them often threatened by fault attacks:
    a single fault during the computation of a cryptographic primitive can lead to a total loss of system security.
    This can have serious consequences, e.g., in safety-critical systems, including bodily harm and catastrophic technical failures.
    However, countermeasures often focus on isolated fault models and high layers of abstraction.
    This leads to a dangerous sense of security, because exploitable faults that are only visible at machine code level might not be covered by countermeasures.

    In this work we present \fsim, a fully automated open source framework for exhaustive fault simulation on binaries of the ubiquitous ARM-M class.
    It allows engineers and analysts to efficiently scan a binary for potential weaknesses against arbitrary combinations of multi-variate fault injections under a large variety of fault models.
    Using \fsim, we demonstrate the power of fully automated fault analysis and the dangerous implications of applying countermeasures without knowledge of physical addresses and offsets.
    We exemplarily analyze two case studies, which are highly relevant for practice: a DFA on AES (cryptographic) and a secure bootloader (non-cryptographic).
    
    Our results show that indeed numerous exploitable faults found by \fsim which occur in the actual implementations are easily missed in manual inspection.
    Crucially, most faults are only visible when taking machine code information, i.e., addresses and offsets, into account.
    Surprisingly, we show that a countermeasure that protects against one type of fault can actually largely increase the vulnerability to other fault models.
    Our work demonstrates the need for countermeasures that, at least in their evaluation, are not restricted to isolated fault models and consider low-level information during the design process.
\end{abstract}

% !TeX root = main.tex

\section{Introduction}
Around 1970, it was discovered that cosmic radiation can induce charge in microelectronics, resulting in potentially faulty behavior of unprotected microcontrollers in avionics.
Since then, securing embedded systems against faults has been a major interest of the safety community.
In the late 1990s,  \textit{fault detection} and \textit{fault prevention} has also moved into the spotlight of the security community.
One of the earliest and best known examples for such a fault attack is the Bellcore attack on CRT-RSA signatures~\cite{boneh1997importance}.
Over the last two decades, numerous generic and specific fault attacks have been proposed~\cite{verbauwhede2011fault} that threaten many if not most embedded systems.
Fortunately, there exist also a multitude of countermeasures against various kinds of faults, which are routinely implemented in security-sensitive applications.

However, practically evaluating the resistance of a system against fault attacks might become a tedious task because of the large parameter space:
Only if physical parameters such as fault intensity, duration, timing etc.\ as well as logical parameters (e.g., location, clock cycle) are correct, a useful fault might appear.
%Further, exhaustively stepping through all combinations might be infeasible due to long intervals in between injections, e.g., through the involved interface to the device or the time required to adjust the fault injection setup.
This led to different approaches of assessing fault attacks on various levels of abstraction.

Evaluating faulty behavior on the lowest design levels produces reliable results~\cite{lu2013laser}, but is only possible if a detailed transistor-, gate-level-, or RTL-description of the hardware is available.
Yet, especially for off-the-shelf microcontrollers such as IoT devices, a detailed low-level description is rarely available.
On the other end of the spectrum, there is recent work that evaluates abstract descriptions of algorithms or implementations with respect to some specific mathematical fault exploitation~\cite{breier2018fault, saha2018expfault}.
However, it is difficult to transfer this outcome to other types of attacks or even non-cryptographic but still security-critical parts of the implementation.
Even more crucial, applying and evaluating countermeasures at high-level code or even at Assembly code with labels can lead to many faults that are missed in actual physical implementations and a false sense of security.
In fact, we later demonstrate the dangerous implications of neglecting low-level information such as physical addresses or offsets since numerous faults are only surfacing in machine code.

In this work, we take a trade-off between the both extremes described above via hardware emulation.
Focusing on the ubiquitous ARM-M architecture, we introduce \fsim, an automated framework for fault simulation.
\fsim supports highly customizable multi-variate fault models without being restricted to cryptographic applications.
Working directly on machine code, it supports all processors of the \arm{6} and \arm{7} families, since emulation is based only on public architecture information.
Thus, it supports all processors of these families out-of-the-box and further can easily be extended with profiled hardware properties of specific chips.
We use our framework to explore the impact of various design parameters and security approaches on fault vulnerability and uncover numerous faults that are easily missed in manual inspection.

\paragraph{Contributions}
This work provides four major contributions:
\begin{itemize}
    \item We present \emu, an efficient emulator for the \arm{6} and \arm{7} architectures, outperforming current state-of-the-art emulators, and designed with fault injection in mind.
        \emu can be restricted to a specific instruction set/architecture and correctly handles faulty Assembly, both are features that no other existing emulator provides.
    \item We present \fsim, a fully automated fault simulator for ARM-M binaries.
        Based on \emu, it exhaustively simulates arbitrary combinations of a variety of customizable fault models on a given binary, including higher-order faults.
    \item In five analyses we use \fsim to explore the impact of several parameters, i.e., compiler optimization, code positioning, well-known countermeasures, higher-order fault injection, and knowledge of undefined behavior on exploitability.
        All analyses are carried out on two case studies that are relevant in practice, one being cryptographic (a DFA on AES) and one non-cryptographic (a secure bootloader).
        Our results highlight the versatility of \fsim and demonstrate that numerous exploitable faults are outside of the engineers control, easily missed in manual inspection, and cannot be easily prevented through countermeasures.
        Crucially, many of these faults are \textit{completetely} invisible on higher level of abstraction, i.e., C code or even Assembly with labels.
    \item We demonstrate that the application of a countermeasure against faults of a certain model can actually severely increase vulnerability against other fault models.
        One might be easily tempted to think that applying a countermeasure at least does not influence vulnerability or even benefits against other attacks.
        Using \fsim, this effect can be easily analyzed in any application.
\end{itemize}

\section*{Availability}
\label{fsim::sec::availability}
The latest version of \emu and \fsim are available open source at \url{https://github.com/emsec/arm-fault-simulator}.
Our results and visualizations as well as the code of our case studies for full reproducibility are available at \url{https://github.com/emsec/arm-fault-simulator-paper-results}.

% !TeX root = main.tex

\section{Background}
\label{fsim::sec::background}

\subsection{Fault Injection, Evaluation, and Countermeasures}
Since its first introduction by Lenstra and Boneh \etal \cite{Lenstra:164524, boneh1997importance}, both scientific and industrial interest in fault injection are constantly rising.
Numerous physical methods to induce a fault were found, e.g., fault injection through voltage~\cite{aumuller2002fault} or clock glitches~\cite{agoyan2010clocks} and via electromagnetic radiation~\cite{quisquater2002eddy}.
Further, laser fault injection was first introduced by Skorobogatov \etal \cite{skorobogatov2002optical} and can be seen as one of the most precise techniques, allowing to target individual transistors.
Likewise, beyond \ac{DFA}~\cite{boneh1997importance}, numerous advanced mathematical attacks were proposed -- its latest prominent variant being Statistical Ineffective Fault Attacks~\cite{dobraunig2018sifa}.
The sheer amount of different types and classes of fault attacks has repeatedly lead to classification works~\cite{bar2006sorcerer,verbauwhede2011fault,karaklajic2013hardware}.
%Here, common properties are derived such as the offered (or required) precision for a fault method or attack respectively.
The most widely used fault types are single bit set, reset, or flip faults as well as faults targeting entire registers, bytes, words, or variables.
Other properties include the distribution of the resulting faulty values (e.g., biased or uniform), the temporal precision, the duration, the abstraction level targeting transistors or values during a protocol etc.

Recently, injecting multiple useful faults at once has been practically proven using a double spot laser station~\cite{selmke2016attack}.
Injecting multiple faults during a single program execution is often referred to as a \textit{multivariate fault} or a \textit{higher-order fault}.
Throughout this work, we will use the term higher-order fault to precisely indicate the number of injected faults, i.e., an $n$-th order fault is comprised of $n$ individual faults.
To avoid confusion with the overarching term \textit{fault} we sometimes employ \textit{fault combination} to capture all individual faults in a higher-order fault where disambiguation would be difficult.

\revise{Considering countermeasures, one option is to address the physical effect itself, either through shielding (e.g., generating the clock internally, metal shielding to hinder EM pulses etc.) or by deploying sensors, e.g., to detect voltage or laser pulses.
However, those usually only counters a limited number of faulting methods.
Instead, fault detection or correction can be easily achieved by introducing some form of redundancy. 
For example, one could verify after creating signature to counter fault attacks on RSA-CRT~\cite{boneh2001importance}.
For symmetric ciphers, this translates to decrypting after encrypting, comparing the result to the original plaintext.
Another option is two encrypt twice and comparing the result.
This can be done either in parallel (in hardware) or successively, storing the result of the first iteration.
Randomizing a delay between the two iterations already leads to a very effective countermeasure as it is unlikely that an identical fault can be injection twice with the correct timing (without triggering fault counters before).
Instead of checking for consistency only once at the end, so-called Concurrent Error Detection~(CED) schemes check at certain points during the encryption, e.g., after each round (cf.\,\cite{DBLP:journals/pieee/BarenghiBKN12,DBLP:journals/jce/GuoMJK15}).
More advanced schemes deploy coding techniques to achieve a certain $n$-bit fault resistance at a lower cost~\cite{aghaie2019impeccable}.
This, however, also leads to the question of what to do if indeed an error is detected.
One approach is counting the number of faults and deleting secret keys when a certain threshold is reached.
Simply omitting the output might not be enough when dealing with ineffective fault analysis \cite{DBLP:conf/ches/Clavier07} or safe-error attacks~\cite{DBLP:journals/tc/YenJ00}.
Its latest variant, Statistical Ineffective Fault Analysis (SIFA)~\cite{DBLP:journals/tches/DobraunigEKMMP18}, requires even more sophisticated countermeasures if deleting the key is unwanted~\cite{DBLP:journals/iacr/BreierKH019,DBLP:journals/tches/DaemenDEGMP20}.
An orthogonal approach is Infective Computation\cite{DBLP:journals/tc/YenJ00}, where a fault will randomize the internal state so that the faulty ciphertext is random as well.
Unfortunately, such countermeasures at least for symmetric schemes did not receive much trust as those were broken repeatedly~\cite{DBLP:conf/fdtc/BattistelloG13}.}

\subsection{Fault Simulation and Assessment}
From an attacker's perspective, a useful fault might only appear if all involved parameters are correct.
Finding working configurations exhaustively, e.g., for location, timing, intensity, etc., might be impossible due to the immense individual parameter space.
This already led to various approaches to improve the search for suitable parameters~\cite{carpi2013glitch,schellenberg2015complexity}.
Instead of trying different parameter sets, simulation techniques were leveraged to speed up fault analysis.
Ironically, laser fault injection itself was originally introduced to simulate heavy ion strikes in electronics that would happen in avionics.
Nowadays, various approaches aim to entirely simulate the effects of targeted faults~\cite{lu2013laser,papadimitriou2016analysis,papadimitriou2015validation,papadimitriou2014multiple}.
Here, one or multiple injections are simulated either on the RTL level or down at the transistor level using SPICE.
For the latter, the resulting faulty behavior is again evaluated solely digital once the fault is latched.
In~\cite{papadimitriou2015validation}, the authors even evaluate layout information with respect to the size of the laser spot.
This all, of course, necessitates many implementation details of the device under test.
Although desired, in the security context such a detailed description is often not available, e.g., in smartcards or multipurpose microcontrollers such as the ARM-platform targeted here.

%The works above stop when generating any faulty output that would be visible to an attacker although this does not always result in exploitable behavior, depending on the application.
On the other hand, there are various approaches that try to find an actual mathematical attack based on some description of a cipher, e.g., using algebraic fault attacks~\cite{gay2019hardware} or classical \ac{DFA}~\cite{breier2018fault} (cf.\,\cite{breier:2019:book} for an overview).
%The work in~\cite{breier2018fault} might seem similar to our setting at first glance.
%However, based on a data flow graph that is extracted from the Assembly, it tries to find spots that pose a suitable distinguisher for \ac{DFA}.
%Thus, it is only applicable to the cryptographic operation itself and not to other security-critical parts of the code.
However, these works are tailored to specific types of mathematical fault analysis and the results usually cannot be transferred to other fault attacks.
Even further, those are only applicable to the cryptographic operation itself and not to other security-critical parts of the code.
For example \cite{breier2018fault} searches for vulnerable spots for \ac{DFA} based on a data flow graph extracted from Assembly.
A similar approach is presented in~\cite{goubet2015efficient}, but instead, the extracted set of equations is fed to an SMT solver to find a distinguisher.
%its focus on DFA is only suitable for the cryptographic operation itself and it only focuses on first-order faults.
Abstracting further, \cite{saha2018expfault, khanna2017xfc} operate on a mathematical description of a cryptographic operation to find a possible attack in the first place.
As we highlight later, this might be a good early indication, but its actual implementation adds numerous additional potential faults that are invisible at a high abstraction level.
%\cite{khairallah2018dfarpa} DFARPA: differential fault attack resistant physical design automation %not important
Riscure sketches with FiSim~\cite{riscure2018fisim}, a proprietary closed-source tool for ARM fault simulation.
Yet, it only focuses on first order attacks and is based on Unicorn (which is inadequate for fault simulation as we discuss in \Section{fsim::sec::emu}).
A similar but bottom-up approach is taken in~\cite{DBLP:conf/cardis/DureuilPCDC15}, although only limited to memory faults, closed source, and again restricted to single faults.

\revise{The authors of \cite{DBLP:conf/trustcom/Given-WilsonJLL17} use bounded model checking to automatically verify whether tailored assertions also hold on a modified binary. To this end, the assertions are added in the source code to be compiled alongside. The binary is then modified according to the fault model. Both the genuine and the faulty binary are transferred back to LLVM-IR to be able to check whether the assertions hold. Unfortunately, this approach has some large performance penalty (2 hours for verifying single-bit opcode faults for code that checks a PIN in a loop). Further, formulating assertions on source level that should also cover possible vulnerabilities on assembly level is hard and error-prone.}
% !TeX root = main.tex

\section{Motivation}
\label{fsim::sec::motivation}
While fault simulation on hardware models would be the most accurate approach, it is very difficult if not impossible to get the hardware models of off-the-shelf processors for analysis.
Thus, we consider fault simulation on the binary level, which provides a good trade-off between coverage and accessibility.
More concretely, we focus on binary-level fault simulation of ARM-M processors, i.e., we simulate faults on the final machine code that is flashed into the processor.
With ARM being the market leader in embedded systems and the ubiquity of ARM-M processors, especially in IoT products, this platform is one of the most relevant for fault attacks.
The specifications of the ARM-M instruction set architectures \arm{6} and \arm{7} are public, so covering the specifications means covering all \acp{IC} of said architecture regardless of vendor/manufacturer.

From a research perspective, we want to demonstrate a crucial problem of current software countermeasures against fault injection: they mainly operate on a high level of abstraction, e.g., on C-code.
Since faults are eventually affecting the final machine code, restricting the analysis to high-level code can result in exploitable faults that are easily missed in both automated and manual inspection.
We show that even a comparatively low-level description like Assembly with labels still hides many exploitable faults, since physical addresses and offsets are still not available.
Our results demonstrate the need for fault detection schemes that take the final machine code into account.
The tools presented here, \emu and \fsim, provide a comprehensive framework that, we believe, can substantially support both scientific research and evaluation of real-world implementations.

% !TeX root = main.tex

\section{\emu}
\label{fsim::sec::emu}
In order to simulate faults on a binary in software, the binary has to be run in an emulator.
For the ARM architecture, several emulators exist, including VisUAL\footnote{\url{https://salmanarif.bitbucket.io/visual/index.html}}, thumbulator\footnote{\url{https://github.com/dwelch67/thumbulator}}, and state-of-the-art QEMU-based Unicorn\footnote{\url{https://www.unicorn-engine.org/}}.
However, purely graphical emulators such as VisUAL are not suitable for automated tools.
Unfortunately, the other existing emulators share a crucial property that make them unsuitable for fault simulation: they do not correctly handle faulty Assembly and are not architecture-restrictable.

These emulators were built to emulate valid ARM Assembly and perfectly provide this functionality.
Developers took shortcuts to optimize efficiency and neglected invalid operand combinations, which is perfectly reasonable for assembler-generated machine code.
However, when faced with invalid machine code as common in the case of faults, these emulators behave incorrectly or simply crash.
For example, a 32-bit instruction may be faulted and now represents two 16-bit instructions.
If dynamically injected at the time of execution, available emulators execute the first 16-bit instruction but skip over the second one since they originally hit a 32-bit instruction.
More subtly, when analyzing an \arm{6} based processor and an instruction is faulted into a valid \arm{7}-only instruction, existing emulators will simply execute the function, even though the analyzed processor is not capable of doing so.
This potentially leads to false positives.

In total, for correct fault simulation, an emulator is needed that is \textit{correct in the presence of faults} and \textit{architecture aware}.
Therefore, we present \emu (pronounced ['$\epsilon$m /ulator/]), our own emulator for \arm{6} and \arm{7} binaries that features both desired properties.
Written from scratch in C++17, it provides a simple but powerful user interface and full control over all simulated aspects of the processor.
Every access to registers or memory can be hooked with custom functions, both before and after access.
Likewise, instruction execution can be hooked before fetch, after decode, and after execute.
This is particularly useful for fault injection since the emulator respects all changes made during the hook functions.
For example, a hook before instruction decode can be used to emulate a fault, e.g., on the instruction fetch register or on the memory bus.
This results in a different (and potentially malformed) instruction being executed.

The architecture of \emu is held simple for maintainability and clarity, consisting mainly of two parts: the instruction decoder and the execution unit.
The instruction decoder is implemented as a large if/else tree based on ARMs machine readable instruction set description, i.e., it precisely mirrors the public specification.
While this results in a lot of code, every instruction decoding has its dedicated place and can easily be located, read, debugged, or tweaked.
The execution unit takes a decoded instruction and updates registers/memory accordingly, firing all registered hooks in the process.
For functional correctness we tested \emu against state-of-the-art emulator Unicorn by comparing CPU and memory states after executing all supported instructions.

In terms of efficiency, \emu outperforms Unicorn.
We compiled both emulators with the same GCC version 7.4.0 and their default release options and let them emulate the same \arm{7} binary on the same PC.
For a binary that computes an AES-128 key schedule followed by 100 encryptions \emu took 155 ms and Unicorn took 333 ms.
For 100 SHA256 computations \emu took 271 ms and Unicorn took 574 ms on the same binary.
The above times were averaged over 50 executions to account for OS scheduling, caches etc.
In total, \emu finished these small-scale tests roughly 2.1 times as fast as Unicorn.

Naturally, an emulator is only an abstraction or approximation of the real hardware.
However, we will show in \Section{fsim::sec::analyses_overview::analysis5} that \emu can easily be extended with profiling data of a real chip in order to be even more accurate for specific hardware.

% !TeX root = main.tex

\section{\fsim}
\label{fsim::sec::armory}
\begin{figure}[h]
    \centering
    \resizebox{\linewidth}{!}{
    \begin{tikzpicture}[x=1cm,y=1cm]

    \node[rectangle, draw, minimum width = 2.6cm] (models) at (-7,6) {\begin{tabular}{c}Fault\\Models\end{tabular}};
    \node[rectangle, draw, minimum width = 2.6cm] (emu) at (-4,6) {\begin{tabular}{c}\emu\\Instance\end{tabular}};
    \node[rectangle, draw, minimum width = 2.6cm] (explmodel) at (-1,6) {\begin{tabular}{c}Exploitability\\Model\end{tabular}};
    \node[rectangle, draw, minimum width = 2.6cm] (endAddr) at (2,6) {\begin{tabular}{c}Halting\\Points\end{tabular}};

    \draw[line width = 1.5] (-8.5,5) rectangle (3.5,-3.5);
    \node[anchor = north west] at (-8.5,5) {\fsim};

    \node[circle,fill=black] (start) at (-4,4.5) {};

    \node[rectangle, draw] (listFaults) at (-4,3.5) {List Injection Points};
    \node[rectangle, draw] (getNext) at (-4,2.5) {Advance to Next Injection Point};

    \node[rectangle, draw] (inject) at (-4,1.5) {Apply Current Fault Model};
    \node[rectangle, draw] (runEmu) at (-4,0.5) {Run Mulator};

    \node[rectangle, draw, double, double distance = 1pt] (rec) at (-4,-1.5) {Recursively Start With Next Fault Model};
    \node[rectangle, draw] (foundFault) at (-2.5,-2.5) {Found Exploitable Fault};

    \node (endHit) at (-2,-0.5) {\textcolor{black!40!green}{Halting Point Hit}};
    \node (noEndHit) at (-6,-0.5) {\textcolor{red!80!black}{No Halting Point Hit}};

    \node[circle,draw=black, minimum width = 0.5cm] (exit) at (2,2.5) {};
    \node[circle,fill=black] at (2,2.5) {};

    \node[rectangle, draw] (eval) at (2,-1.5) {\begin{tabular}{c}Evaluate\\Exploitability\\Model\end{tabular}};

    \tikzstyle{arrow}=[->, shorten > = 0.1em, shorten < = 0.1em]

    \draw[arrow] (emu) -- ($(emu) - (0,1)$);
    \draw[arrow] (models) -- ($(models) - (0,1)$);
    \draw[arrow] (endAddr) -- ($(endAddr) - (0,1)$);
    \draw[arrow] (explmodel) -- ($(explmodel) - (0,1)$);

    \draw[arrow] (start) -- (listFaults);
    \draw[arrow] (listFaults) -- (getNext);
    \draw[arrow] (getNext) -- (exit)  node[pos = .5,below] {\textcolor{red!80!black}{no more remaining}} ;
    \draw[arrow] (getNext) -- (inject);
    \draw[arrow] (inject) -- (runEmu);
    \draw[arrow] (runEmu) -- (endHit) -| (eval);
    \draw[arrow]  (eval) |- (foundFault)  node[pos=.73,below] {\textcolor{black!40!green}{\checkmark}};
    \draw[arrow,shorten < = 0em] (eval) -- (rec)  node[pos = .5,below] {\textcolor{red!80!black}{\crossout}};

    \draw[arrow] (runEmu) -- (noEndHit) -- (noEndHit |- rec.north);

    \draw[arrow] (-6.5,-1.75) -- (-6.5,-3) -| ($(getNext) - (4.25,0.5)$) |- ($(getNext) +(0,0.5)$);
    \draw[arrow,-,shorten > = 0em] (foundFault) |- ($(-6.5,-2.5) - (0,0.5)$);
    \end{tikzpicture}}
    \caption{Simplified workflow of \fsim}
    \label{fsim::overview}
\end{figure}
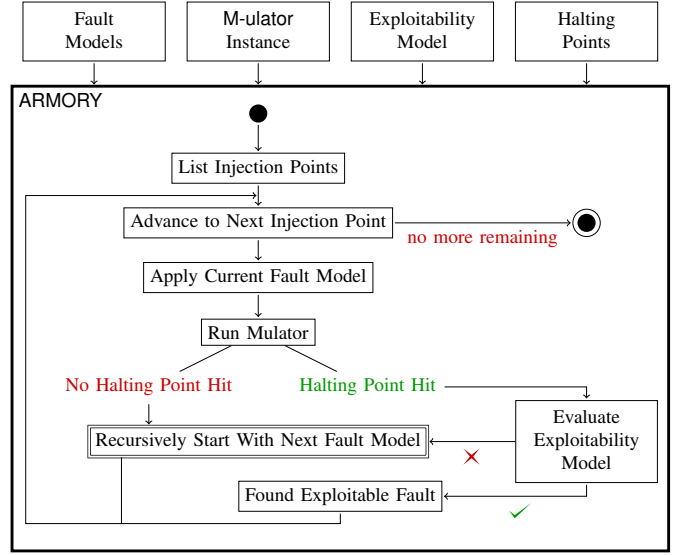

In this section we present \fsim, our automated fault injection tool.
\fsim is built on \emu and is capable of exhaustively injecting arbitrary combinations of arbitrary fault models.
The general workflow is illustrated in \Figure{fsim::overview}.
For the moment we focus on explaining the inputs and general functionality.
We dive into the inner workings later in \Section{fsim::sec::armory::strategy}.
In the end, in \Section{fsim::sec::armory::realworld}, we provide a detailed discussion on applying \fsim in real-world scenarios.

\subsection{\revise{Terminology}}
\label{fsim::sec::armory::terminology}

\revise{
    As described in \Section{fsim::sec::emu}, \emu is a instruction-accurate emulator, i.e., it is not cycle accurate.
    This is due to different pipeline architectures across the \arm{6} and \arm{7} families and a lack of cycle descriptions in the official documentation.
    Hence, \fsim provides details on the timing of fault injection down to instruction-level granularity.
    Therefore we use the index of an instruction in the instruction sequence of the program as our notion of \textit{time}, i.e., if a fault has to be injected at time 5 this means it has to be injected on the 5$^\text{th}$ executed instruction.
}

\revise{
    \fsim exhaustively simulates faults on a binary, i.e., it applies the given fault models to every possible \enquote{position} or \enquote{location}.
    For instance, a instruction bit flip fault will be applied to all 16 or 32 bits of each instruction and the resulting effects are observed.
    A register byte fault would be applied to all four bytes of each 32-bit register every time it is used.
    In the following we collectively refer to these points where a fault model is applied as \textit{injection points}.
    We again would like to stress that \fsim will exhaustively simulate the effects of faults on all possible injection points in the whole binary.
}

\revise{
    Eventually, \fsim has to decide whether a fault is exploitable.
    This is covered via an \textit{exploitability model}.
    Such a model can cover simple cases such as ``incorrect output observed'', but also supports way more complex cases, e.g., a critical function was executed or a specific algebraic relation between output bytes is met.
}

\subsection{Inputs}
\label{fsim::sec::armory::inputs}
\revise{
    A user has to provide four inputs to \fsim: (1) an \emu instance, (2) a set of fault models to inject, (3) an exploitability model, and (4) a set of addresses at which the exploitability shall be evaluated, so called \textit{halting points}.
}

The first input is an \emu instance, loaded with the binary to test.
The user can prepare the instance with arbitrary hooks and emulate as many instructions as he wants in order to bring the emulator into the desired initial state.
Of course, all instructions executed until this point are fault-free.

\revise{
The second input is a set of fault models for \fsim to test.
Details on the supported fault models will be given in the \Section{fsim::sec::armory::models}.
To the best of our knowledge, we support the largest set of fault models found in literature among all simulation-based approaches.
More fault models can be added easily without the need to change \fsim.
}

\revise{
The third input is an exploitability model, since, eventually, \fsim has to decide whether a fault injection led to exploitable behavior.
While prior work oftentimes simply regards incorrect output as exploitable, depending on the application this may be oversimplified or does not capture an adversary's objective.
Hence, the user has to provide an exploitability model, i.e., a set of conditions which, when met by injecting a fault, regards the fault as exploitable.
Such an exploitability model enables far more flexible vulnerability assessment than in previous simulation-based approaches, as it can cover everything from simple output mismatch tests, over more complex arithmetic checks, to even functionality-checks, e.g., whether specific (sensitive) code was executed.
}

\revise{
The fourth and final input is a set of \textit{halting points}.
Since exploitability typically surfaces only after numerous instructions have been executed, e.g., after a ciphertext is output, it would be unreasonable and inefficient to verify the exploitability model after each instruction.
Hence, these halting points define addresses of instructions where the exploitability model is evaluated.
}

The user can also set the maximum number of fault models applied at the same time, set the maximum number of utilized CPU cores, specify address ranges in which no fault should be applied, and give a timeout in form of the maximum number of executed instructions to avoid infinite loops.
For the sake of simplicity these features were not shown in \Figure{fsim::overview}.

\subsection{Outputs}
\label{fsim::sec::armory::outputs}
Given its inputs, \fsim exhaustively simulates all supplied models in all possible combinations on all available instructions/registers without further user interaction and reports all exploitable fault combinations.
The reported fault combinations contain all information necessary to understand and replicate the faulty behavior\revise{, i.e., injection time, affected instruction or register, affected bit/byte, etc}.

\fsim also comes with a fault tracer, which allows to verify a specific fault combination and, as the name suggests, to trace the effects of faults instruction-by-instruction.

Since the modeled behavior is purely based on publicly available instruction specifications by ARM, all detected faults are guaranteed to work on real hardware, given that the attacker is able to precisely inject the required fault models.
Hence there is a guaranteed 0\% false positive rate.
However, note that emulation is just an abstraction of the real hardware.
Therefore, \fsim cannot capture all faults that were possible on real hardware.
For example, \fsim does not know about internal buffer registers etc, hence it cannot simulate faults on these entities.
Yet, if such internal behavior is known to the user, it can be added.
\revise{We provide more details on applying \fsim in real-world scenarios in \Section{fsim::sec::armory::realworld}.}

\subsection{Fault Models}
\label{fsim::sec::armory::models}
As previously mentioned, \fsim is capable of applying arbitrary fault models to the execution.
The current implementation supports arbitrary faults on instructions and faults on registers.
\revise{
    Note that, while this selection may sound fairly limited, instruction-level and register-level faults are sufficient to model a wide range of real-world fault effects as we describe in \Section{fsim::sec::armory::realworld}.
}
However, because of the modular design of \fsim, other classes of fault models can be added with ease.
Every fault model is highly configurable, from its effect, over its duration, to specific filters, e.g., to skip faulting the \texttt{LR} register or to skip branch instructions.
Note that each model may also result in multiple injection points on a single instruction or register, e.g., a register byte-fault would be applied 4 times.
A complete example of a fault model would be a \texttt{permanent, 4-iteration, only-on-R0, register-byte-set}-fault.

\revise{
    We stress that \fsim cannot decide whether a fault model is actually feasible in the real world.
    A detailed discussion on \fsim in real-world scenarios is given in \Section{fsim::sec::armory::realworld}.
}

\paragraph{\revise{Instruction Fault Specifics}}
Instruction faults can be either \textit{permanent} or \textit{transient}.
A permanent fault is injected at the beginning of simulation and affects the instruction every time it is executed, while a transient fault is only active for a single instruction execution.
Permanent instruction faults can be used to model, e.g., faults in the program memory, while transient instruction faults can model, e.g., faults on the instruction decoder register or instruction bus.

\paragraph{\revise{Register Fault Specifics}}
Register faults can be either \textit{permanent}, \textit{active until overwrite}, or \textit{transient}.
A permanent fault affects the register value at the beginning of simulation and on all subsequent writes, an until-overwrite fault affects a register value once but subsequent (over-)writes to the register will remove the fault's effect, while a transient fault is only active for a single instruction execution.
While register faults mostly model laser faults on register banks, a transient register fault can also model, e.g., a fault on the transistors that drive a register's output wires.

\subsection{Optimized Fault Simulation Strategy}
\label{fsim::sec::armory::strategy}
During simulation, \fsim exhaustively applies the given fault models at every possible injection point, i.e., location or position where the fault model is applicable, cf.~\Section{fsim::sec::armory::terminology}.
However, the numerous possibilities for fault models pose a challenge for simulation of higher-order faults.
A single fault might completely change the control flow, thus rendering static evaluation of injection points incorrect.
Therefore, injection points are determined dynamically at runtime during a single fault-free simulation.
We will first describe fault simulation strategy in general, before explaining the details of higher-order fault simulation (cf.~\Figure{fsim::overview})
%Our explanations follow the depiction of \fsim in \Figure{fsim::overview}.

\paragraph{Fault Simulation Strategy}
%\color{blue}
First, \fsim needs to find injection points for faulting, i.e., all locations where the given fault models are applicable.
Therefore, the emulator is ran without injecting faults, i.e., a dry run, until the binary finishes or the supplied timeout is reached.
This provides the sequence of executed instructions and used registers.
This sequence directly holds all injection points in sorted order.
For example, regarding instruction skip faults, every instruction of the sequence is a single injection point, while for a single-bit register fault every register of the sequence marks 32 injection points.
Fault simulation is then performed as follows:

Starting from the beginning, the emulator is ran until right before the next injection point and its current state is stored as backup.
Now the fault model is applied and the emulator continues.
If an halting point is reached, the exploitability model is checked to decide whether the currently injected fault combination is exploitable.
In that case, the current fault is appended to the list of exploitable faults, otherwise the fault is regarded as not exploitable.
Likewise, a fault is not exploitable if no halting point was reached until simulation stops or \emu encounters an invalid instruction, e.g., an undefined opcode, invalid operands, etc.
The emulator's state is now restored to the backup right before the fault was injected and the emulator is advanced to the next injection point
This backup/restore mechanic removes the necessity of having to start from the beginning for each fault, heavily reducing emulation time (more details in \Section{fsim::sec::armory::optimizations}).
Then the injection process repeats.
Note that, in case a fault model is applied multiple times to one instruction or register, \emu does not need to be advanced at all in between injections.
%\color{black}

\paragraph{Higher-Order Fault Simulation}
While the above simulation technique works well for isolated first-order faults, higher-order (multivariate) faults are also of interest.
However, in this scenario not only is the order of injected faults important but also their immediate effects:
One fault might result in a change in control flow which enables a whole range of new faults.

\revise{
    Intuitively, the complexity of simulating a list of fault models is $\mathcal{O}(n^m)$ with $n$ being the number of injection points and $m$ being the number of fault models.
}

Every fault model starts with the aforementioned dry run to identify injection points.
Then, fault simulation is performed with the current model for the current fault combination as explained above.
If the fault combination was found to be exploitable, the next iteration is analyzed as usual.
However, if the fault combination is not exploitable but still produced valid machine code, \fsim starts recursively from the current state with the next fault model.
Hence, the next model starts with an \emu where a specific fault combination has already been applied, i.e., its subsequent dry-run captures the effects of all previously induced faults, and then starts to inject its own faults on all newly identified injection points.

This way, the given list of fault models is applied to all possible combinations of instructions/registers.
However, the order of fault models is also important, i.e., the outcome might be different depending on which is injected first.
\revise{
In addition, unfiltered higher-order fault simulation results in a lot of redundant results:
if, e.g., a first-order fault is known to be exploitable, adding a second fault to the execution is oftentimes exploitable as well but also a waste of computation time.
On the other hand, the additional faults can also invalidate the exploitable effect of the first fault, e.g., by crashing the chip or triggering other fault detection mechanisms.
However, naturally, lower-order faults are much more severe than higher-order faults since their attack complexity is considerably lower.
In other words, invalidating a vulnerability by injecting additional faults does not fix the vulnerability, hence it is reasonable to optimize for the least-effort-requiring vulnerability.
Therefore, we filter out such redundant faults in our fault simulation workflow:
\fsim performs fault injection simulation for all permutations of the input fault models while keeping already found fault combinations in mind.
If at any point a fault combination would be injected for which a subset of included faults already was found to be exploitable, this fault combination is skipped.
This saves an entire recursion every time, i.e., roughly $\mathcal{O}(n^{(m-i)})$ operations with $i$ being the index of the current fault model.
}

Since applying all fault models in all permutations is computationally quite complex, we applied several optimizations to keep performance high.
First, permanent faults are never permuted since they are all injected at the beginning.
Hence, for permanent faults, it is enough to only consider all subsets, while for the non-permanent faults, indeed all subsets and their permutations have to be considered.
On top of that, when applying a specific fault model multiple times, permutations of this specific model do not matter as well.
\hyperref[fsim::lst::permutations]{Listing~\ref*{fsim::lst::permutations}} shows an example of all tested fault model combinations for two different permanent fault models and two different non-permanent models when a maximum of three fault models at the same time is set.

While the actual impact of this optimization depends on the binary and the employed fault models, in our tests we achieved an improvement in execution time of up to two orders of magnitude.
However, since the actual saving depends on the models used and the faults detected in the iterations, we cannot give a better complexity expression than upper bounding the number of permutations pessimistically with $m!$.

Therefore, the overall complexity of higher-order fault analysis is roughly $\mathcal{O}(n^{m} \cdot m!)$, although several internal optimizations allow us to regard this as a very pessimistic.
\revise{
    Note that this complexity results from the problem of exhaustive simulation itself and is not due to the architecture of \fsim.
    We would like to note that actual simulation time cannot be accurately approximated by such formula since the eventual effects of faults on the binary play a major role during simulation:
    invalid Assembly can be regarded as not exploitable immediately, while a change in a loop counter may result in several hundred more instruction executions before a decision on exploitability can be made.
}

\begin{figure}[!t]
    \centering
    \begin{minipage}[t]{0.3\linewidth}
        \centerline{\texttt{P1}}
        \centerline{\texttt{N1}}
        \centerline{\texttt{N2}}
    \end{minipage}
    \hfill
    \begin{minipage}[t]{0.3\linewidth}
        \centerline{\texttt{P1, P2}}
        \centerline{\texttt{P1, N1}}
        \centerline{\texttt{P1, N2}}
        \centerline{\texttt{P2, N1}}
        \centerline{\texttt{P2, N2}}
        \centerline{\texttt{N1, N2}}
        \centerline{\texttt{N2, N1}}
    \end{minipage}
    \hfill
    \begin{minipage}[t]{0.3\linewidth}
        \centerline{\texttt{P1, P2, N1}}
        \centerline{\texttt{P1, P2, N2}}
        \centerline{\texttt{P1, N1, N2}}
        \centerline{\texttt{P1, N2, N1}}
        \centerline{\texttt{P2, N1, N2}}
        \centerline{\texttt{P2, N2, N1}}
    \end{minipage}\\
    \captionof{lstlisting}{Tested fault model combinations up to length 3 for two permanent fault models \texttt{P1} and \texttt{P2} and two non-permanent models \texttt{N1} and \texttt{N2}.}
    \label{fsim::lst::permutations}
\end{figure}

\subsection{Further Optimizations}
\label{fsim::sec::armory::optimizations}
\revise{
Since especially higher-order fault simulation is computationally taxing, we employed several additional optimizations to increase performance where possible.
Note that these optimizations are generic strategies, tailored to our scenario, and automatically also improve first-order fault simulation.
}

\paragraph{\emu Backups}
As described in \Section{fsim::sec::armory::strategy}, the emulator's state is often saved and restored in order to reduce emulation time.
This backup process includes saving/restoring all registers, CPU-state information, and RAM contents.
Especially the latter is a notable bottleneck if the entire memory is copied all the time.
Therefore, we went with the following strategy:
on the initial backup, the whole RAM content is copied.
\revise{
Then we make use of \emu's before-memory-write hook and remember the start and end addresses of the memory range that will be overwritten.
}
We specifically remember two ranges, one near the stack pointer and one near the start of RAM, since these are the most common spots of RAM access.
Upon the next backup, only the modified RAM-ranges have to be updated from the emulator and likewise, when restoring a backup, only the stored ranges have to be copied back to the emulator's memory.
This optimized backup system again improved performance by an order of magnitude.

\paragraph{Efficient Multicore Support}
\fsim can automatically utilize all available cores by splitting the workload of the first fault model over all available CPU cores. % (if not configured otherwise).
Therefore, as long as the list of injection points for the first fault model has more entries than cores are available, efficiency scales linearly with available CPU cores.
However, work is not simply split in equal parts, but rather processed in a queue that every thread fetches from.
Simulation time varies heavily depending on the effect of a specific fault, especially if multiple models are to be tested.
\revise{
    For instance, while one fault might immediately result in invalid Assembly, another fault might result in an infinite loop which would be simulated until reaching the timeout.
}
Thus, a simple split of the work in equal parts would result in some threads finishing way faster than others.
In turn this would not utilize all CPU cores optimally.
The queue solution results in every core processing the next injection point that is available until, in total, all injection points are processed.
This way, cores will only idle when all injection points have already been processed or are currently assigned to other cores.

We noted in \Section{fsim::sec::armory::strategy} that the order of permanent fault models does not matter.
However, permanent register faults can only be applied to the 17 available registers in ARM (\texttt{R0}-\texttt{R15} and \texttt{xPSR}).
Hence, if a permanent register fault model is the first in the list of fault models, only a maximum of 17 cores can be utilized.
Therefore, the permanent fault models are automatically sorted to apply instruction faults first, if available.
Note that this has no impact on the output, but benefits multithreading.

\subsection{\revise{Applying \fsim in Real-World Scenarios}}
\label{fsim::sec::armory::realworld}
%\color{blue}
\fsim has several useful applications, however the main focus is always on automating parts of the process of an analyst.
The analyst can, for example, be part of an evaluation lab's certification team that analyzes an external product or an in-house security engineer that tries to detect vulnerabilities in a product.
In a typical application, an analyst would first gather fault models which he regards as realistic or applicable, given the device under inspection, and then invoke \fsim to test (parts of) the binary for vulnerability with respect to application-specific criteria which he formalizes in the exploitability model.

\paragraph{Injecting Detected Faults into Real-World \acp{IC}}
Note that \fsim does not decide whether a fault model is actually feasible in the real world.
As for all simulation-based approaches, an analyst has to decide, based on his own equipment or known state-of-the-art, which fault models seem reasonable.
Likewise, because of its abstraction level, \fsim does not take noise during fault injection into account.
However, this is strictly in line with previous work on fault injection simulation:
only the \textbf{effects} of faults are analyzed via fault models, the actual physical injection is not in scope of binary-level simulators.

\paragraph{Supported Fault Effects}
For our analyses, we use an exemplary set of models that are commonly found in literature.
Since our fault models are based on instruction-level and register-level faults, it might seem that they are insufficient to model a lot of real-world fault effects.
However, in the following we exemplarily outline that our fault models indeed cover a wide variety of common real-world fault effects:
\begin{itemize}
    \item \textbf{Instruction skipping:} by replacing the current instruction with a \texttt{NOP}, instruction skip faults can be modeled.
    \item \textbf{Faults on flash (program memory):} any fault on program memory is equivalent to faulting instructions loaded from specific addresses.
    \item \textbf{Faults on RAM (data memory):} any fault on data memory can be modeled via a register fault after loading or before storing.
    \item \textbf{Faulting operand registers:} bit/byte faults on instructions easily change the used registers of an instruction. More targeted modification is possible as well.
    \item \textbf{Faulting addresses:} faults on the immediate offsets of instructions or register faults on address registers can change the source/destination of a memory access.
    \item \textbf{Diverting control flow:} faults on instructions can turn many of them into branches, change the used branch condition, or simply change the branch offset.
    Register faults on the status register(s) can also change the outcome of a previous condition evaluation and by faulting the program counter register, the processor immediately branches to a different location in the binary.
    \item \textbf{Instruction replacement:} direct replacements of specific instructions or instruction types is supported via the instruction fault models.
        Note that, e.g., single-bit faults on instructions also commonly result in different instructions if the opcode encoding is faulted.
\end{itemize}
All above real-world fault effects are covered by \fsim.
Note that this list is \textbf{not} exhaustive, there are many more effects that our current fault models can represent.
However, this list shows that \fsim is indeed not restricted in its workings.

During physical fault injection and simulation alike, invalid instructions may be loaded by the processor, e.g., invalid register operands or undefined operations.
This is caught by \emu, since it correctly reports cases of undefined behavior etc.
\fsim can then mark the currently tested fault as not exploitable as a real chip would either crash/reset or execute undefined behavior which is beyond the scope of general simulation.
However, we show in our fifth experiment in \Section{fsim::sec::analyses_overview::analysis5}, that \emu can be easily extended with profiled behavior of a specific chip to even handle cases where simulation would otherwise have to be aborted due to undefined behavior.

\paragraph{Guaranteed Correctness}
Since \emu is based solely on public architectural information and skips any cases of uncertainty, e.g., whenever undefined behavior is encountered, \fsim will never generate false positives.
However, while \fsim can simulate all kinds of instruction and register faults, it is up to the analyst to decide which fault models are realistic on the target platform.
In other words, if one is able to reliably inject a given fault model, all exploitable faults that \fsim reports for this model are guaranteed to work on real hardware.

\paragraph{Timing Information}
A major difficulty in fault analysis is finding the point in time when a specific fault has to be applied.
While \emu is not cycle-accurate as this information is implementation dependent (cf.~\Section{fsim::sec::armory::terminology}), every reported fault contains timing information as the exact position in the sequence of executed instructions.
This helps analysts to vastly narrow down the search space in time for applying a specific fault.
Note that this is also helpful when only rather imprecise faults are possible in practice:
If \fsim indicates numerous exploitable faults at a certain period in time, fault injection in general seems to be more successful in that period.
This information can then be leveraged by analysts to run a constrained brute-force attack instead of faulting over the whole program execution.

\paragraph{Downsides of Abstraction}
As mentioned in \Section{fsim::sec::emu}, emulation is always an abstraction.
Hence, \fsim can miss exploitable faults that, for example, are enabled by internal registers or a platform's specific handling of undefined behavior.
Commonly detailed design descriptions of the device under inspection is not available, hence the employed level of abstraction is still the most detailed representation that is generically available.

Note that \fsim is specifically engineered and optimized for this scenario.
If for a specific design hardware internals are known, integrating this knowledge into \fsim is not straightforward and would require changes to several distinct parts of \fsim and \emu.

Furthermore, given that an analyst provides an incorrect exploitability mode, i.e., a model that does not capture all conditions under which an adversary could compromise security, \fsim will not detect these vulnerabilities.
However, this is a general issue of fault assessment techniques.
As a mitigation, the analyst could employ highly generic exploitability models such as ``unexpected output observed'', but this might result in many faults being reported as exploitable, which would not yield any useful information to an adversary.

%\color{black}

% !TeX root = main.tex
\newpage
\section{Case Studies}
\label{fsim::sec::casestudies}
In order to demonstrate that not taking final machine code into account leaves exploitable faults that are easily missed in manual analysis (cf.\ \Section{fsim::sec::motivation}), we apply \fsim to two case studies in various settings.
The first case study focuses on the non-cryptographic scenario of secure boot while the second case study targets \ac{DFA} on AES.
These are two well known scenarios in research on fault attacks and also relevant in practice.
Both case studies are used in five analyses, each exploring a different setting.
Our analyses also highlight the applicability and versatility of \fsim for both cryptographic and non-cryptographic scenarios.

In this section we describe the setup of the two case studies.
The code of all case studies, the complete outputs of our experiments, and all additional data are available \hyperref[fsim::sec::availability]{on GitHub}.
We implemented both case studies in C and compiled them to ARM Assembly using ARM-GCC v6.3.1.

\subsection{Case Study: Secure Boot}
\label{fsim::sec::casestudies::secureboot}
Many embedded systems feature an update mechanism in order to deliver improved firmware to already deployed devices.
However, especially in the \ac{IoT}, developers want to ensure that only their own firmware can be executed on the device.
Therefore, developers employ a \textit{secure boot mechanism}~\cite{riscure2018fisim}:
The device is equipped with a small ROM that contains a fixed bootloader that is burned-in during production.
Upon startup, this bootloader fetches the current firmware from an external memory and verifies that it stems from an authorized source.
If the firmware is authorized, it is copied to RAM and executed, otherwise the system is shut down.
The verification process is typically implemented via cryptographic signatures and can also be implemented in several layers.

\paragraph{Attacker Target}
If an attacker manages to skip or trick firmware verification, he can execute arbitrary code on the device, potentially reading out sensitive data.
Thus, the goal of the attacker is to introduce faults such that the bootloader executes an unauthorized firmware.
Note that we explicitly do not try to break the employed cryptographic mechanism, hence classify this scenario as non-cryptographic.

\paragraph{Case Study Implementation}
For the sake of simplicity, we implemented the firmware verification by computing an unpadded SHA256 checksum and comparing it to an expected value.
The firmware in our case study has exactly 128 bytes (externally supplied, unknown to the compiler), i.e., the SHA will process the firmware in two blocks internally.
Note that these simplifications do not make our results less representative, since we never target the cryptographic computations themselves.
Still, we uncover interesting exploitable faults that are independent of the concrete verification mechanism as visible in our results.
During fault simulation, the supplied firmware never matches the expected SHA value, i.e., without exploitable faults the bootloader will always abort.
The overall functionality can be summarized by the following pythonic pseudocode:
\newpage
\begin{lstlisting}[numbers=none]
def main(firmware, expected_hash):
    hash = sha256(firmware)
    for i from 0 to 31:
        if hash[i] != expected_hash[i]:
            report_error()
            return
    execute_firmware()
\end{lstlisting}
The code of the \texttt{sha256} function is part of the binary.

\paragraph{Exploitability Model}
A fault combination is regarded as exploitable if it leads to execution of the \texttt{execute\_firmware()} function although the hash values do not match.
Hence, exploitability in this case study is highly control-flow-dependent.

\subsection{Case Study: AES}
\label{fsim::sec::casestudies::aes}
Our second case study focuses on \ac{DFA} of the AES cipher. %differential fault analysis
Note that several variants of \ac{DFA} on AES exist, depending on the type of fault and faulting precision.
W.l.o.g., we focus on a specific byte-fault:
If an attacker is able to obtain faulty ciphertexts where a fault affects a single byte after the MixColumns operation of round 7 and before the final SubBytes operation, he is able to drastically reduce the search space of the key.
Typically, only two ciphertext-faultytext-pairs per key byte are required to recover the whole key.
For more details on \ac{DFA} of AES we refer to~\cite{dusart2003differential,piret2003differential}.

\paragraph{Attacker Target}
Following the above description, the attacker's goal is to inject faults which affect a single byte of the AES state in the specified period.
Note that it is not required to control the exact outcome of the fault, it is only required to affect a single byte.

\paragraph{Case Study Implementation}
\revise{As the base of our implementation we chose the well-known open-source \enquote{Tiny-AES}\footnote{Tiny-AES on GitHub: \url{https://github.com/kokke/tiny-AES-c}}, a full AES implementation for embedded devices which is optimized for small code size.}
The overall functionality can be summarized by the following pythonic pseudocode:
\begin{lstlisting}[numbers=none]
def main(plaintxt, masterkey):
    rkeys = key_schedule(masterkey)
    ciphertext = encrypt(plaintxt, rkeys)
    report_done()
\end{lstlisting}
Since faults that apply before MixColumns of round 7 are not exploitable in the discussed \ac{DFA} variant, \fsim is invoked with an \emu instance which has already been run until that point.

\paragraph{Exploitability Model}
A fault combination is regarded as exploitable if the \texttt{report\_done()} function is executed and the resulting ciphertext has exactly a single faulty byte in the state somewhere after MixColumns of round 7 and before the final SubBytes operation.
In \fsim this is checked by computing the AES backwards on the resulting ciphertext and comparing the intermediate state with the known intermediate values of a fault-free encryption.
Exploitability in this case study is highly data-dependent.

\revise{
    Note that this model is a \ac{DFA}-tailored version of the simple ``wrong output is received'' model, which would report a large amount of false positives that cannot be exploited in a \ac{DFA}.
}

% !TeX root = main.tex

\section{Overview on Performed Analyses}
The initial motivation of our work is that several exploitable faults are easily missed in both automated and manual inspection, if low-level information, e.g., physical addresses and offsets, are neglected in the analysis (cf.\ \Section{fsim::sec::motivation}).
We use \fsim to perform five analyses of our two case studies to demonstrate the severity and truth of that statement.

In the following we present the parameters and approaches we analyze.
The results are discussed subsequently in \Section{fsim::sec::results}.
If not stated otherwise, all analyses were performed on the compiler-generated \arm{7} Assembly code of our two case studies for multiple optimization targets.
In order to compactly describe a specific test case, we use square brackets, e.g., [AES-DFA, \textit{no countermeasure}, \texttt{O3}].

We configured the 24 fault models shown in \Table{fsim::tab::eval::model_legend} for \fsim.
In the following we refer to the fault models by their number.
Note that, while reliable targeted bit flips have never been reported through physical fault injection, the fault model is still reasonable for faults that are only applied to a static value, i.e., they capture the isolated change of a bit.
Thus, for the permanent register faults, we had to split the bit-flip model into bit-set and bit-clear since the register values are faulted anytime they change.
Register fault models were configured to be applied to registers \texttt{R0}--\texttt{R12}, \texttt{LR} and \texttt{xPSR}.
\begin{table}[tb]
	\begin{footnotesize}
    \begin{minipage}[t]{0.3\linewidth}
        \centering
        \begin{tabular}{rrc}
        	\toprule
        	                       \multicolumn{3}{c}{Instruction Faults}                        \\ \midrule
        	                    \multicolumn{2}{c}{Model}                     & \multicolumn{1}{c}{\#} \\ \cmidrule(r){1-2}\cmidrule(l){3-3}
        	\multirow{4}{*}{\rotatebox[origin=c]{90}{Permanent}} &       Skip &       \model{1}        \\
        	                                                     &   Byte-Set &       \model{2}        \\
        	                                                     & Byte-Clear &       \model{3}        \\
        	                                                     &   Bit-Flip &       \model{4}        \\ \arrayrulecolor{lightgray}\midrule\arrayrulecolor{black}
        	\multirow{4}{*}{\rotatebox[origin=c]{90}{Transient}} &       Skip &       \model{5}        \\
        	                                                     &   Byte-Set &       \model{6}        \\
        	                                                     & Byte-Clear &       \model{7}        \\
        	                                                     &   Bit-Flip &       \model{8}        \\ \bottomrule
        \end{tabular}
    \end{minipage}
   \hfill
    \begin{minipage}[t]{0.3\linewidth}
        \centering
        \begin{tabular}{rrc}
        	\toprule
        	                                  \multicolumn{3}{c}{Register Faults}                                    \\ \midrule
        	                              \multicolumn{2}{c}{Model}                               & \multicolumn{1}{c}{\#} \\ \cmidrule(r){1-2}\cmidrule(l){3-3}
        	                    \multirow{6}{*}{\rotatebox[origin=c]{90}{Permanent}} &      Clear &       \model{9}        \\
        	                                                                         &       Fill &       \model{10}       \\
        	                                                                         &   Byte-Set &       \model{11}       \\
        	                                                                         & Byte-Clear &       \model{12}       \\
        	                                                                         &    Bit-Set &       \model{13}       \\
        	                                                                         &  Bit-Clear &       \model{14}       \\ \bottomrule
        \end{tabular}
    \end{minipage}
\hfill
	\begin{minipage}[t]{0.3\linewidth}
	\centering
	\begin{tabular}{rrc}
		\toprule
        	                                  \multicolumn{3}{c}{Register Faults}                                    \\ \midrule
\multicolumn{2}{c}{Model}                               & \multicolumn{1}{c}{\#} \\ \cmidrule(r){1-2}\cmidrule(l){3-3}
\multirow{5}{*}{\rotatebox[origin=c]{90}{\shortstack{Until Overwr.}}} &      Clear &       \model{15}       \\
&       Fill &       \model{16}       \\
&   Byte-Set &       \model{17}       \\
& Byte-Clear &       \model{18}       \\
&   Bit-Flip &       \model{19}       \\ \arrayrulecolor{lightgray}\midrule\arrayrulecolor{black}
\multirow{5}{*}{\rotatebox[origin=c]{90}{Transient}} &      Clear &       \model{20}       \\
&       Fill &       \model{21}       \\
&   Byte-Set &       \model{22}       \\
& Byte-Clear &       \model{23}       \\
&   Bit-Flip &       \model{24}       \\ \bottomrule
	\end{tabular}
	\end{minipage}
\end{footnotesize}
    \caption{Fault models used in the analyses}
    \label{fsim::tab::eval::model_legend}
\end{table}
% ################################################################################
% ################################################################################
% ################################################################################

\subsection{Analysis I -- Influence of Compiler Optimization}
\label{fsim::sec::analyses_overview::analysis1}
Our first analysis forms the base for all other analyses.
We investigate the general fault-exploitability of the two case studies and take a look at the impact of different compiler optimization levels.

We compiled both case studies using the four optimization targets \texttt{O1}, \texttt{O2}, \texttt{O3}, and \texttt{Os} and ran \fsim on the compiler-generated assembly code.
\fsim tested all 24 fault models in first order, i.e., injecting a single fault in one execution, for a total of 192 tested combinations.

% ################################################################################
% ################################################################################
% ################################################################################

\subsection{Analysis II -- Influence of Code Positioning}
\label{fsim::sec::analyses_overview::analysis2}
In our second analysis, we examine the impact of relative code positioning.
Note that the relative position of code blocks is not necessarily visible in high-level languages.
We show that especially control flow-related faults heavily depend on the position within the binary, i.e., whether a certain address can be reached through a fault.

Again we analyzed both case studies with the four compiler optimization levels from \hyperref[fsim::sec::analyses_overview::analysis1]{Analysis I}.
There, the SHA256 code and the AES key schedule were located at the lowest addresses (starting at \texttt{0x8000} in our case) followed by the remaining code of the respective case study.
For this analysis we swapped these code blocks, leaving the SHA256 or the AES key schedule at higher address spaces.
In both cases, \fsim applied all 24 fault models as first order faults

% ################################################################################
% ################################################################################
% ################################################################################

\subsection{Analysis III -- Influence of Countermeasures}
\label{fsim::sec::analyses_overview::analysis3}
Our third analysis focuses on the impact of basic countermeasures on the \textit{overall} number of exploitable faults.
Specific countermeasures mostly target a single class of faults and are evaluated by their success in detecting or preventing faults of this specific fault class.
However, the overall impact on exploitability, i.e., including faults of other models, is often unknown.

We selected two straightforward countermeasures, namely (1) instruction replacement~\cite{moro2014formal} and (2) control flow checking through signatures~\cite{oh2002control}.
These are two well known principal techniques on which many follow-up countermeasures are based.
They can also be jointly applied without mutually interfering.

\paragraph{Instruction Replacement}
Instruction replacement is a \textit{fault prevention} technique and aims at protecting against instruction skip faults by duplicating all instructions or splitting them into an equivalent set of instructions if direct duplication is not possible.
The technique was thoroughly analyzed and formally proven to protect against first-order instruction skip faults by Moro \etal \cite{moro2014formal}.
We implemented the countermeasure following their descriptions.

\paragraph{Control Flow Check}
Control flow checking through signatures is a \textit{fault detection} technique by Oh \etal \cite{oh2002control} and aims at protecting control flow, i.e., it aims to ensure that basic blocks are executed in the intended order.
It reserves a register as the Global Signature Register (GSR) and assigns a random unique signature to each basic block.
The general premise is that a basic block can verify its predecessor through the GSR.
When transitioning from basic block $A$ to basic block $B$, the GSR is updated as $GSR := GSR \text{ xor } (ID_A \text{ xor } ID_B)$, and then compared to the expected $ID_B$.
If the results match, the control flow was correct, and the GSR now holds the ID of the new basic block.
Otherwise, a fault was detected.
The authors acknowledge that several kinds of control flow faults still cannot be detected by their technique, e.g., branches within a basic block.

\paragraph{Test Cases}
We ran \fsim on the compiler-generated Assembly code of both case studies for all optimization targets from \hyperref[fsim::sec::analyses_overview::analysis1]{Analysis I}, first without any protection, then with each technique in isolation, and finally with both techniques combined.
This leaves us with 32 combinations which were tested for exploitable faults.
We applied all 24 fault models in first order.

% ################################################################################
% ################################################################################
% ################################################################################

\subsection{Analysis IV -- Injecting Higher-Order Faults}
\label{fsim::sec::analyses_overview::analysis4}
In the fourth analysis, we examine higher-order fault injection.
Most countermeasures are dedicated to first-order faults, i.e., a single fault in the whole computation.
However, depending on the physical setup, an attacker may be able to inject more than just one fault, i.e., at multiple positions in time and/or at different positions on the device.

We selected all configurations of the AES-DFA case study from \hyperref[fsim::sec::analyses_overview::analysis3]{Analysis III}, i.e., with and without countermeasures using the four optimization options.
The secure boot case study was only analyzed without countermeasures, since the code was too large to exhaustively simulate higher-order faults in a reasonable timeframe.

Fault simulation was performed up to the second order with all fault models that did not target single bits.
%, i.e., excluding models \model{4}, \model{8}, \model{13}, \model{14}, \model{19}, and \model{24}.
%Up to two faults of the same model were tested in one execution, i.e., \fsim injected second-order faults.

% ################################################################################
% ################################################################################
% ################################################################################

\subsection{Analysis V -- Profiling Undefined Behavior}
\label{fsim::sec::analyses_overview::analysis5}
In the last analysis, we focus on the abstraction penalty through emulation.
Naturally, an emulator will not be able to find all faults that are possible in a real-world attack.
For example, whenever \emu hits an encoding or a parameter combination which is specified as undefined or unpredictable, it has to abort emulation.
However, by profiling a specific processor and implementing the actual behavior of these cases into \emu, \fsim can assess the real processor more accurately.
We will collectively address all undefined, unpredictable, etc.\ cases as \textit{undefined behavior} in this context.

We selected two platforms, namely (1) the Infineon XMC 2Go featuring the ARM Cortex-M0 based XMC1100 processor (\arm{6}) and (2) the STM32F4 Discovery board featuring the ARM Cortex-M4 based STM32F4VGT processor (\arm{7}).
For the small \arm{6} architecture, only eight instructions can actually trigger undefined behavior.
We profiled all of them by manually manipulating a compiled test binary and observing chip behavior with a hardware debugger on-chip.
For the \arm{7} architecture, the number of undefined cases is too large to check manually, so we focused on the same instructions as in the \arm{6} setting.

% !TeX root = main.tex

\section{Results}
\label{fsim::sec::results}
In this section we present the results of the five studies described above and  discuss their implications.

% ################################################################################
% ################################################################################
% ################################################################################

\subsection{Results of Analysis I -- Influence of Compiler Optimization}
\label{fsim::sec::results::analysis1}

\begin{table*}[t]
    \centering
    \begin{adjustbox}{max width=\textwidth}
        \begin{tabular}{crrrrrrrrrrrrrrrrrrrrrrrrrrrrrrr}
    \toprule
    Opt. &  \multicolumn{1}{c}{Time} &  \multicolumn{8}{c}{Instruction Fault Models} & \multicolumn{14}{c}{Register Fault Models} \\ \cmidrule(r){1-1} \cmidrule(lr){2-2} \cmidrule(lr){3-10}  \cmidrule(l){11-24} 
    & &  \multicolumn{4}{c}{Permanent} &  \multicolumn{4}{c}{Transient} &  \multicolumn{4}{c}{Permanent} &  \multicolumn{5}{c}{Until Overwrite} & \multicolumn{5}{c}{Transient} \\
    &       &  \multicolumn{1}{c}{\model{1}} &  \multicolumn{1}{c}{\model{2}} &   \multicolumn{1}{c}{\model{3}} &   \multicolumn{1}{c}{\model{4}} &   \multicolumn{1}{c}{\model{5}} &   \multicolumn{1}{c}{\model{6}} &    \multicolumn{1}{c}{\model{7}} &    \multicolumn{1}{c}{\model{8}} & \multicolumn{1}{c}{\model{11}} & \multicolumn{1}{c}{\model{12}} & \multicolumn{1}{c}{\model{13}} & \multicolumn{1}{c}{\model{14}} &  \multicolumn{1}{c}{\model{15}} &  \multicolumn{1}{c}{\model{16}} &   \multicolumn{1}{c}{\model{17}} &  \multicolumn{1}{c}{\model{18}} &   \multicolumn{1}{c}{\model{19}} &  \multicolumn{1}{c}{\model{20}} &  \multicolumn{1}{c}{\model{21}} &   \multicolumn{1}{c}{\model{22}} &   \multicolumn{1}{c}{\model{23}} &    \multicolumn{1}{c}{\model{24}}  \\ 
    
    \cmidrule(lr){3-6} \cmidrule(lr){7-10}  \cmidrule(lr){11-14}  \cmidrule(lr){15-19}   \cmidrule(l){20-24}
    \midrule
    \multicolumn{24}{c}{Analysis I: Secure Boot} \\ \cmidrule(r){1-1} \cmidrule(lr){2-2} \cmidrule(lr){3-6} \cmidrule(lr){7-10} \cmidrule(lr){11-14} \cmidrule(lr){15-19} \cmidrule(l){20-24}
    O1 & 10489 & 0 & 0 & 0 & 2 & 0 & 0 & 0 & 15 & 1 & 0 & 2 & 1 & 0 & 0 & 0 & 0 & 2 & 0 & 0 & 0 & 0 & 2\\
    O2 & 10742 & 1 & 1 & 0 & 7 & 1 & 0 & 0 & 4 & 1 & 0 & 2 & 0 & 0 & 0 & 0 & 0 & 2 & 0 & 0 & 0 & 0 & 2\\
    O3 & 6892 & 1 & 1 & 0 & 8 & 1 & 0 & 0 & 6 & 1 & 0 & 1 & 0 & 0 & 0 & 0 & 0 & 1 & 0 & 0 & 0 & 0 & 1\\
    Os & 10825 & 1 & 1 & 0 & 2 & 0 & 0 & 0 & 1 & 1 & 1 & 1 & 0 & 0 & 0 & 0 & 0 & 0 & 0 & 0 & 0 & 0 & 0\\
    \midrule
    \multicolumn{24}{c}{Analysis III: Secure Boot with Instruction Replacement Countermeasure}\\ \cmidrule(r){1-1} \cmidrule(lr){2-2} \cmidrule(lr){3-6} \cmidrule(lr){7-10} \cmidrule(lr){11-14} \cmidrule(lr){15-19} \cmidrule(l){20-24}
    O1 & 31261 & \textcolor{gray}{0} & \textcolor{red}{+1} & \textcolor{gray}{0} & \textcolor{red}{+6} & \textcolor{gray}{0} & \textcolor{red}{+1} & \textcolor{gray}{0} & \textcolor{red}{+247} & \textcolor{gray}{0} & \textcolor{gray}{0} & \textcolor{green!60!black}{-1} & \textcolor{green!60!black}{-1} & \textcolor{gray}{0} & \textcolor{gray}{0} & \textcolor{gray}{0} & \textcolor{gray}{0} & \textcolor{green!60!black}{-2} & \textcolor{gray}{0} & \textcolor{gray}{0} & \textcolor{gray}{0} & \textcolor{gray}{0} & \textcolor{green!60!black}{-2}\\
    O2 & 29328 & \textcolor{green!60!black}{-1} & \textcolor{red}{+1} & \textcolor{gray}{0} & \textcolor{red}{+12} & \textcolor{green!60!black}{-1} & \textcolor{red}{+1} & \textcolor{gray}{0} & \textcolor{red}{+83} & \textcolor{gray}{0} & \textcolor{gray}{0} & \textcolor{green!60!black}{-1} & \textcolor{gray}{0} & \textcolor{gray}{0} & \textcolor{gray}{0} & \textcolor{gray}{0} & \textcolor{gray}{0} & \textcolor{green!60!black}{-2} & \textcolor{gray}{0} & \textcolor{gray}{0} & \textcolor{gray}{0} & \textcolor{gray}{0} & \textcolor{green!60!black}{-2}\\
    O3 & 19255 & \textcolor{green!60!black}{-1} & \textcolor{red}{+2} & \textcolor{gray}{0} & \textcolor{red}{+21} & \textcolor{green!60!black}{-1} & \textcolor{red}{+2} & \textcolor{gray}{0} & \textcolor{red}{+363} & \textcolor{gray}{0} & \textcolor{gray}{0} & \textcolor{gray}{0} & \textcolor{gray}{0} & \textcolor{gray}{0} & \textcolor{gray}{0} & \textcolor{gray}{0} & \textcolor{red}{+2} & \textcolor{green!60!black}{-1} & \textcolor{gray}{0} & \textcolor{gray}{0} & \textcolor{gray}{0} & \textcolor{gray}{0} & \textcolor{green!60!black}{-1}\\
    Os & 33267 & \textcolor{green!60!black}{-1} & \textcolor{red}{+2} & \textcolor{gray}{0} & \textcolor{red}{+10} & \textcolor{gray}{0} & \textcolor{red}{+2} & \textcolor{gray}{0} & \textcolor{red}{+212} & \textcolor{gray}{0} & \textcolor{gray}{0} & \textcolor{gray}{0} & \textcolor{gray}{0} & \textcolor{gray}{0} & \textcolor{gray}{0} & \textcolor{gray}{0} & \textcolor{gray}{0} & \textcolor{gray}{0} & \textcolor{gray}{0} & \textcolor{gray}{0} & \textcolor{gray}{0} & \textcolor{gray}{0} & \textcolor{gray}{0}\\
    \midrule
    \multicolumn{24}{c}{Analysis III: Secure Boot with Control Flow Checking Countermeasure}\\ \cmidrule(r){1-1} \cmidrule(lr){2-2} \cmidrule(lr){3-6} \cmidrule(lr){7-10} \cmidrule(lr){11-14} \cmidrule(lr){15-19} \cmidrule(l){20-24}
    O1 & 22258 & \textcolor{gray}{0} & \textcolor{gray}{0} & \textcolor{gray}{0} & \textcolor{green!60!black}{-2} & \textcolor{gray}{0} & \textcolor{gray}{0} & \textcolor{gray}{0} & \textcolor{green!60!black}{-15} & \textcolor{gray}{0} & \textcolor{gray}{0} & \textcolor{green!60!black}{-1} & \textcolor{green!60!black}{-1} & \textcolor{gray}{0} & \textcolor{gray}{0} & \textcolor{gray}{0} & \textcolor{gray}{0} & \textcolor{green!60!black}{-2} & \textcolor{gray}{0} & \textcolor{gray}{0} & \textcolor{gray}{0} & \textcolor{gray}{0} & \textcolor{green!60!black}{-2}\\
    O2 & 29476 & \textcolor{green!60!black}{-1} & \textcolor{green!60!black}{-1} & \textcolor{gray}{0} & \textcolor{green!60!black}{-4} & \textcolor{green!60!black}{-1} & \textcolor{gray}{0} & \textcolor{gray}{0} & \textcolor{green!60!black}{-4} & \textcolor{gray}{0} & \textcolor{gray}{0} & \textcolor{green!60!black}{-1} & \textcolor{gray}{0} & \textcolor{gray}{0} & \textcolor{gray}{0} & \textcolor{gray}{0} & \textcolor{gray}{0} & \textcolor{green!60!black}{-2} & \textcolor{gray}{0} & \textcolor{gray}{0} & \textcolor{gray}{0} & \textcolor{gray}{0} & \textcolor{green!60!black}{-2}\\
    O3 & 10155 & \textcolor{green!60!black}{-1} & \textcolor{green!60!black}{-1} & \textcolor{gray}{0} & \textcolor{green!60!black}{-5} & \textcolor{green!60!black}{-1} & \textcolor{gray}{0} & \textcolor{gray}{0} & \textcolor{green!60!black}{-6} & \textcolor{gray}{0} & \textcolor{gray}{0} & \textcolor{gray}{0} & \textcolor{gray}{0} & \textcolor{gray}{0} & \textcolor{gray}{0} & \textcolor{gray}{0} & \textcolor{gray}{0} & \textcolor{green!60!black}{-1} & \textcolor{gray}{0} & \textcolor{gray}{0} & \textcolor{gray}{0} & \textcolor{gray}{0} & \textcolor{green!60!black}{-1}\\
    Os & 21737 & \textcolor{gray}{0} & \textcolor{green!60!black}{-1} & \textcolor{gray}{0} & \textcolor{green!60!black}{-1} & \textcolor{gray}{0} & \textcolor{gray}{0} & \textcolor{gray}{0} & \textcolor{green!60!black}{-1} & \textcolor{gray}{0} & \textcolor{green!60!black}{-1} & \textcolor{gray}{0} & \textcolor{gray}{0} & \textcolor{gray}{0} & \textcolor{gray}{0} & \textcolor{gray}{0} & \textcolor{gray}{0} & \textcolor{gray}{0} & \textcolor{gray}{0} & \textcolor{gray}{0} & \textcolor{gray}{0} & \textcolor{gray}{0} & \textcolor{gray}{0}\\
    \midrule
    \multicolumn{24}{c}{Analysis III: Secure Boot with both Countermeasures}\\ \cmidrule(r){1-1} \cmidrule(lr){2-2} \cmidrule(lr){3-6} \cmidrule(lr){7-10} \cmidrule(lr){11-14} \cmidrule(lr){15-19} \cmidrule(l){20-24}
    O1 & 58245 & \textcolor{gray}{0} & \textcolor{gray}{0} & \textcolor{gray}{0} & \textcolor{green!60!black}{-2} & \textcolor{gray}{0} & \textcolor{gray}{0} & \textcolor{gray}{0} & \textcolor{green!60!black}{-15} & \textcolor{gray}{0} & \textcolor{gray}{0} & \textcolor{green!60!black}{-1} & \textcolor{green!60!black}{-1} & \textcolor{gray}{0} & \textcolor{gray}{0} & \textcolor{gray}{0} & \textcolor{gray}{0} & \textcolor{green!60!black}{-2} & \textcolor{gray}{0} & \textcolor{gray}{0} & \textcolor{gray}{0} & \textcolor{gray}{0} & \textcolor{green!60!black}{-2}\\
    O2 & 74632 & \textcolor{green!60!black}{-1} & \textcolor{green!60!black}{-1} & \textcolor{gray}{0} & \textcolor{green!60!black}{-7} & \textcolor{green!60!black}{-1} & \textcolor{gray}{0} & \textcolor{gray}{0} & \textcolor{green!60!black}{-4} & \textcolor{gray}{0} & \textcolor{gray}{0} & \textcolor{green!60!black}{-1} & \textcolor{gray}{0} & \textcolor{gray}{0} & \textcolor{gray}{0} & \textcolor{gray}{0} & \textcolor{gray}{0} & \textcolor{green!60!black}{-2} & \textcolor{gray}{0} & \textcolor{gray}{0} & \textcolor{gray}{0} & \textcolor{gray}{0} & \textcolor{green!60!black}{-2}\\
    O3 & 27615 & \textcolor{green!60!black}{-1} & \textcolor{green!60!black}{-1} & \textcolor{gray}{0} & \textcolor{green!60!black}{-8} & \textcolor{green!60!black}{-1} & \textcolor{gray}{0} & \textcolor{gray}{0} & \textcolor{green!60!black}{-6} & \textcolor{gray}{0} & \textcolor{gray}{0} & \textcolor{gray}{0} & \textcolor{gray}{0} & \textcolor{gray}{0} & \textcolor{gray}{0} & \textcolor{gray}{0} & \textcolor{gray}{0} & \textcolor{green!60!black}{-1} & \textcolor{gray}{0} & \textcolor{gray}{0} & \textcolor{gray}{0} & \textcolor{gray}{0} & \textcolor{green!60!black}{-1}\\
    Os & 57313 & \textcolor{green!60!black}{-1} & \textcolor{green!60!black}{-1} & \textcolor{gray}{0} & \textcolor{green!60!black}{-2} & \textcolor{gray}{0} & \textcolor{gray}{0} & \textcolor{gray}{0} & \textcolor{green!60!black}{-1} & \textcolor{gray}{0} & \textcolor{green!60!black}{-1} & \textcolor{gray}{0} & \textcolor{gray}{0} & \textcolor{gray}{0} & \textcolor{gray}{0} & \textcolor{gray}{0} & \textcolor{gray}{0} & \textcolor{gray}{0} & \textcolor{gray}{0} & \textcolor{gray}{0} & \textcolor{gray}{0} & \textcolor{gray}{0} & \textcolor{gray}{0}\\
    \midrule
    \midrule
    \multicolumn{24}{c}{Analysis I: AES-DFA} \\ \cmidrule(r){1-1} \cmidrule(lr){2-2} \cmidrule(lr){3-6} \cmidrule(lr){7-10} \cmidrule(lr){11-14} \cmidrule(lr){15-19} \cmidrule(l){20-24}
    O1 & 1320 & 24 & 15 & 25 & 282 & 568 & 136 & 948 & 8082 & 0 & 0 & 0 & 0 & 610 & 636 & 812 & 648 & 7452 & 621 & 717 & 2090 & 943 & 18010\\
    O2 & 1311 & 24 & 24 & 24 & 215 & 566 & 138 & 959 & 7002 & 0 & 0 & 1 & 0 & 570 & 619 & 705 & 608 & 6656 & 633 & 729 & 2112 & 922 & 17990\\
    O3 & 874 & 73 & 31 & 120 & 920 & 503 & 116 & 971 & 6311 & 0 & 0 & 1 & 40 & 568 & 597 & 751 & 618 & 8899 & 533 & 559 & 1279 & 695 & 12887\\
    Os & 1382 & 24 & 16 & 24 & 240 & 590 & 163 & 1096 & 7908 & 0 & 0 & 0 & 0 & 610 & 636 & 908 & 678 & 7997 & 629 & 724 & 2298 & 938 & 18283\\
    \midrule
    \multicolumn{24}{c}{Analysis III: AES-DFA with Instruction Replacement Countermeasure} \\ \cmidrule(r){1-1} \cmidrule(lr){2-2} \cmidrule(lr){3-6} \cmidrule(lr){7-10} \cmidrule(lr){11-14} \cmidrule(lr){15-19} \cmidrule(l){20-24}
    O1 & 3848 & \textcolor{green!60!black}{-24} & \textcolor{red}{+5} & \textcolor{red}{+1} & \textcolor{red}{+34} & \textcolor{green!60!black}{-568} & \textcolor{red}{+263} & \textcolor{red}{+464} & \textcolor{red}{+2978} & \textcolor{gray}{0} & \textcolor{gray}{0} & \textcolor{gray}{0} & \textcolor{gray}{0} & \textcolor{red}{+1048} & \textcolor{red}{+1130} & \textcolor{red}{+1453} & \textcolor{red}{+1232} & \textcolor{red}{+13398} & \textcolor{red}{+256} & \textcolor{red}{+214} & \textcolor{green!60!black}{-718} & \textcolor{red}{+317} & \textcolor{green!60!black}{-4550}\\
    O2 & 3877 & \textcolor{green!60!black}{-24} & \textcolor{green!60!black}{-4} & \textcolor{red}{+30} & \textcolor{red}{+90} & \textcolor{green!60!black}{-566} & \textcolor{red}{+291} & \textcolor{red}{+595} & \textcolor{red}{+4024} & \textcolor{gray}{0} & \textcolor{gray}{0} & \textcolor{gray}{0} & \textcolor{gray}{0} & \textcolor{red}{+1008} & \textcolor{red}{+1113} & \textcolor{red}{+1693} & \textcolor{red}{+1254} & \textcolor{red}{+14297} & \textcolor{red}{+256} & \textcolor{red}{+214} & \textcolor{green!60!black}{-779} & \textcolor{red}{+334} & \textcolor{green!60!black}{-4713}\\
    O3 & 2291 & \textcolor{green!60!black}{-73} & \textcolor{green!60!black}{-7} & \textcolor{red}{+46} & \textcolor{red}{+233} & \textcolor{green!60!black}{-503} & \textcolor{red}{+124} & \textcolor{red}{+475} & \textcolor{red}{+4848} & \textcolor{gray}{0} & \textcolor{gray}{0} & \textcolor{green!60!black}{-1} & \textcolor{green!60!black}{-2} & \textcolor{red}{+949} & \textcolor{red}{+1008} & \textcolor{red}{+1393} & \textcolor{red}{+1013} & \textcolor{red}{+15240} & \textcolor{red}{+192} & \textcolor{red}{+229} & \textcolor{green!60!black}{-298} & \textcolor{red}{+207} & \textcolor{green!60!black}{-2403}\\
    Os & 4007 & \textcolor{green!60!black}{-24} & \textcolor{green!60!black}{-15} & \textcolor{red}{+2} & \textcolor{red}{+80} & \textcolor{green!60!black}{-590} & \textcolor{red}{+212} & \textcolor{red}{+430} & \textcolor{red}{+2691} & \textcolor{gray}{0} & \textcolor{gray}{0} & \textcolor{gray}{0} & \textcolor{gray}{0} & \textcolor{red}{+1080} & \textcolor{red}{+1162} & \textcolor{red}{+1439} & \textcolor{red}{+1292} & \textcolor{red}{+14563} & \textcolor{red}{+264} & \textcolor{red}{+223} & \textcolor{green!60!black}{-891} & \textcolor{red}{+338} & \textcolor{green!60!black}{-4656}\\
    \midrule
    \multicolumn{24}{c}{Analysis III: AES-DFA with Control Flow Checking Countermeasure} \\ \cmidrule(r){1-1} \cmidrule(lr){2-2} \cmidrule(lr){3-6} \cmidrule(lr){7-10} \cmidrule(lr){11-14} \cmidrule(lr){15-19} \cmidrule(l){20-24}
    O1 & 3074 & \textcolor{gray}{0} & \textcolor{red}{+21} & \textcolor{red}{+39} & \textcolor{red}{+215} & \textcolor{red}{+4} & \textcolor{red}{+139} & \textcolor{red}{+269} & \textcolor{red}{+1103} & \textcolor{gray}{0} & \textcolor{gray}{0} & \textcolor{gray}{0} & \textcolor{gray}{0} & \textcolor{gray}{0} & \textcolor{gray}{0} & \textcolor{red}{+82} & \textcolor{red}{+32} & \textcolor{red}{+704} & \textcolor{gray}{0} & \textcolor{gray}{0} & \textcolor{green!60!black}{-11} & \textcolor{green!60!black}{-9} & \textcolor{red}{+104}\\
    O2 & 3054 & \textcolor{gray}{0} & \textcolor{red}{+12} & \textcolor{red}{+37} & \textcolor{red}{+262} & \textcolor{red}{+1} & \textcolor{red}{+1} & \textcolor{red}{+268} & \textcolor{red}{+1197} & \textcolor{gray}{0} & \textcolor{gray}{0} & \textcolor{green!60!black}{-1} & \textcolor{red}{+1} & \textcolor{gray}{0} & \textcolor{green!60!black}{-1} & \textcolor{red}{+107} & \textcolor{red}{+34} & \textcolor{red}{+727} & \textcolor{gray}{0} & \textcolor{gray}{0} & \textcolor{gray}{0} & \textcolor{red}{+6} & \textcolor{green!60!black}{-92}\\
    O3 & 1060 & \textcolor{red}{+6} & \textcolor{red}{+9} & \textcolor{green!60!black}{-2} & \textcolor{red}{+78} & \textcolor{red}{+111} & \textcolor{red}{+14} & \textcolor{red}{+232} & \textcolor{red}{+1166} & \textcolor{gray}{0} & \textcolor{gray}{0} & \textcolor{red}{+8} & \textcolor{red}{+3} & \textcolor{red}{+28} & \textcolor{red}{+38} & \textcolor{red}{+80} & \textcolor{gray}{0} & \textcolor{red}{+698} & \textcolor{red}{+36} & \textcolor{red}{+32} & \textcolor{red}{+140} & \textcolor{red}{+59} & \textcolor{red}{+1290}\\
    Os & 3114 & \textcolor{gray}{0} & \textcolor{green!60!black}{-16} & \textcolor{red}{+1} & \textcolor{green!60!black}{-49} & \textcolor{red}{+38} & \textcolor{green!60!black}{-46} & \textcolor{red}{+56} & \textcolor{red}{+413} & \textcolor{gray}{0} & \textcolor{gray}{0} & \textcolor{gray}{0} & \textcolor{gray}{0} & \textcolor{red}{+14} & \textcolor{red}{+17} & \textcolor{green!60!black}{-6} & \textcolor{red}{+14} & \textcolor{red}{+650} & \textcolor{red}{+10} & \textcolor{red}{+7} & \textcolor{green!60!black}{-106} & \textcolor{green!60!black}{-8} & \textcolor{red}{+814}\\
    \midrule
    \multicolumn{24}{c}{Analysis III: AES-DFA with both Countermeasures} \\ \cmidrule(r){1-1} \cmidrule(lr){2-2} \cmidrule(lr){3-6} \cmidrule(lr){7-10} \cmidrule(lr){11-14} \cmidrule(lr){15-19} \cmidrule(l){20-24}
    O1 & 7963 & \textcolor{green!60!black}{-24} & \textcolor{green!60!black}{-3} & \textcolor{red}{+25} & \textcolor{red}{+112} & \textcolor{green!60!black}{-568} & \textcolor{red}{+254} & \textcolor{red}{+368} & \textcolor{red}{+3440} & \textcolor{gray}{0} & \textcolor{gray}{0} & \textcolor{gray}{0} & \textcolor{gray}{0} & \textcolor{red}{+1048} & \textcolor{red}{+1130} & \textcolor{red}{+1637} & \textcolor{red}{+1294} & \textcolor{red}{+14770} & \textcolor{red}{+256} & \textcolor{red}{+214} & \textcolor{green!60!black}{-729} & \textcolor{red}{+317} & \textcolor{green!60!black}{-4550}\\
    O2 & 7899 & \textcolor{green!60!black}{-24} & \textcolor{green!60!black}{-12} & \textcolor{red}{+26} & \textcolor{red}{+204} & \textcolor{green!60!black}{-566} & \textcolor{red}{+229} & \textcolor{red}{+557} & \textcolor{red}{+3913} & \textcolor{gray}{0} & \textcolor{gray}{0} & \textcolor{green!60!black}{-1} & \textcolor{red}{+1} & \textcolor{red}{+1008} & \textcolor{red}{+1111} & \textcolor{red}{+1863} & \textcolor{red}{+1258} & \textcolor{red}{+14449} & \textcolor{red}{+256} & \textcolor{red}{+213} & \textcolor{green!60!black}{-759} & \textcolor{red}{+334} & \textcolor{green!60!black}{-4697}\\
    O3 & 2638 & \textcolor{green!60!black}{-73} & \textcolor{red}{+1} & \textcolor{red}{+52} & \textcolor{red}{+306} & \textcolor{green!60!black}{-503} & \textcolor{red}{+125} & \textcolor{red}{+1021} & \textcolor{red}{+4917} & \textcolor{gray}{0} & \textcolor{gray}{0} & \textcolor{red}{+8} & \textcolor{red}{+8} & \textcolor{red}{+975} & \textcolor{red}{+1064} & \textcolor{red}{+1593} & \textcolor{red}{+969} & \textcolor{red}{+15719} & \textcolor{red}{+206} & \textcolor{red}{+252} & \textcolor{green!60!black}{-255} & \textcolor{red}{+221} & \textcolor{green!60!black}{-2118}\\
    Os & 8073 & \textcolor{green!60!black}{-24} & \textcolor{green!60!black}{-16} & \textcolor{red}{+1} & \textcolor{red}{+18} & \textcolor{green!60!black}{-590} & \textcolor{red}{+237} & \textcolor{red}{+401} & \textcolor{red}{+3044} & \textcolor{gray}{0} & \textcolor{gray}{0} & \textcolor{gray}{0} & \textcolor{gray}{0} & \textcolor{red}{+1104} & \textcolor{red}{+1196} & \textcolor{red}{+1749} & \textcolor{red}{+1316} & \textcolor{red}{+15525} & \textcolor{red}{+278} & \textcolor{red}{+239} & \textcolor{green!60!black}{-861} & \textcolor{red}{+352} & \textcolor{green!60!black}{-4415}\\
    \bottomrule
\end{tabular}

    \end{adjustbox}
    \caption{Results of Analyses I and III: Number of exploitable faults by fault model (cf.\,\Table{fsim::tab::eval::model_legend}). Models \model{9} and \model{10} are excluded since they never resulted in exploitable faults. Colored values show the increase of exploitable faults relative to the respective unprotected base case.}
    \label{fsim::tab::experiment1and3::results}
\end{table*}
\fsim took between 0.5 and 22 seconds for each configuration on a single PC utilizing 56 cores.
\Table{fsim::tab::experiment1and3::results} shows the combined results of this analysis and Analysis III.
For each case study and each configuration, we list the time (in number of executed instructions) from the start of the case study until the bootloader detects the modified firmware or the AES returns its ciphertext, followed by the number of exploitable faults for all 24 fault models.
Note that execution times for the AES-DFA case study are relatively low since we start fault simulation after the MixColumns step of the seventh round.
We focus on the results of Analysis I in this section.

The first thing to notice is that the AES-DFA case study shows much more exploitable faults than the secure boot case study.
This originates from the type of faults that are required for a ``successful'' exploitation for the two case studies (cf.\ \Section{fsim::sec::casestudies}):
Exploitability of the secure boot case study is control-flow-dependent and faults that meaningfully change the control flow to circumvent checksum verification are rare.
On the other hand, exploitability of the AES-DFA case study is data-dependent with few constraints.
This enables almost any fault that has a very localized effect to be exploitable.
Notably, permanent register faults (models \model{9}-\model{14}) seem to have very little effect in the AES-DFA case study in general.
%\todo{folgender Satz ist mir nicht klar}\todo{input Falk: das könnte natürlich zu anderen echt mächtigen Fehlern führen, vielleicht does not fit our chosen variant of DFA}
%This is reasonable, since they induce faulty behavior whenever the target register is used, hence rendering the resulting ciphertext useless for the chosen variant of \ac{DFA}.
This is reasonable, since the chosen variant of \ac{DFA} relies on localized data faults, while permanent register faults may affect the processed data multiple times.
However, for the secure boot case study, exploitability with these faults is comparable to the other models.
Depending on how the final hash comparison is implemented, a permanent fault on a register may set all results to ``equals'', effectively circumventing the verification process.
In general, most exploitable faults in the secure boot case study changed a branch target address or turned an instruction into a branch.

\paragraph{General Observations}
The degree of freedom for a successful attack and the type of exploitability, i.e., data-driven or control-flow-driven, of a case study have a major impact on fault-vulnerability.
While naturally with a higher degree of freedom an attacker has more space to work with, it is interesting to see the huge difference in our two case studies.
In terms of compiler optimizations, they definitely have a noticeable impact on exploitability.
However, in our case studies, no consistent ``best optimization'' can be made out, i.e., an engineer cannot simply select a specific optimization level and expect better fault resistance.

% ################################################################################
% ################################################################################
% ################################################################################

\subsection{Results of Analysis II -- Influence of Code Positioning}
\label{fsim::sec::results::analysis2}

\begin{table*}[t]
    \centering
    \begin{adjustbox}{max width=0.85\textwidth}
        \begin{tabular}{crrrrrrrrrrrrrrrrrrrrrrrrrrrrrrrr}
    \toprule
    Opt. &  \multicolumn{8}{c}{Instruction Fault Models} & \multicolumn{16}{c}{Register Fault Models} \\ \cmidrule(r){1-1} \cmidrule(lr){2-9} \cmidrule(l){10-25} 
    &  \multicolumn{4}{c}{Permanent} &  \multicolumn{4}{c}{Transient} &  \multicolumn{6}{c}{Permanent} &  \multicolumn{5}{c}{Until Overwrite} & \multicolumn{5}{c}{Transient} \\
       &  \multicolumn{1}{c}{\model{1}} &  \multicolumn{1}{c}{\model{2}} &   \multicolumn{1}{c}{\model{3}} &   \multicolumn{1}{c}{\model{4}} &   \multicolumn{1}{c}{\model{5}} &   \multicolumn{1}{c}{\model{6}} &    \multicolumn{1}{c}{\model{7}} &    \multicolumn{1}{c}{\model{8}} & \multicolumn{1}{c}{\model{9}} & \multicolumn{1}{c}{\model{10}} & \multicolumn{1}{c}{\model{11}} & \multicolumn{1}{c}{\model{12}} & \multicolumn{1}{c}{\model{13}} & \multicolumn{1}{c}{\model{14}} &  \multicolumn{1}{c}{\model{15}} &  \multicolumn{1}{c}{\model{16}} &   \multicolumn{1}{c}{\model{17}} &  \multicolumn{1}{c}{\model{18}} &   \multicolumn{1}{c}{\model{19}} &  \multicolumn{1}{c}{\model{20}} &  \multicolumn{1}{c}{\model{21}} &   \multicolumn{1}{c}{\model{22}} &   \multicolumn{1}{c}{\model{23}} &    \multicolumn{1}{c}{\model{24}}  \\  
       \cmidrule(lr){2-5} \cmidrule(lr){6-9} \cmidrule(lr){10-15} \cmidrule(lr){16-20} \cmidrule(l){21-25}
    \midrule
    \multicolumn{25}{c}{Analysis II: Secure Boot} \\ \cmidrule(r){1-1} \cmidrule(lr){2-5} \cmidrule(lr){6-9} \cmidrule(lr){10-15} \cmidrule(lr){16-20} \cmidrule(l){21-25} 
    O1 & \textcolor{gray}{0} & \textcolor{gray}{0} & \textcolor{gray}{0} & \textcolor{gray}{0} & \textcolor{gray}{0} & \textcolor{gray}{0} & \textcolor{gray}{0} & -13 & \textcolor{gray}{0} & \textcolor{gray}{0} & \textcolor{gray}{0} & \textcolor{gray}{0} & \textcolor{gray}{0} & \textcolor{gray}{0} & \textcolor{gray}{0} & \textcolor{gray}{0} & \textcolor{gray}{0} & \textcolor{gray}{0} & \textcolor{gray}{0} & \textcolor{gray}{0} & \textcolor{gray}{0} & \textcolor{gray}{0} & \textcolor{gray}{0} & \textcolor{gray}{0}\\
    O2 & \textcolor{gray}{0} & \textcolor{gray}{0} & +1 & +4 & \textcolor{gray}{0} & \textcolor{gray}{0} & +1 & +4 & \textcolor{gray}{0} & \textcolor{gray}{0} & \textcolor{gray}{0} & \textcolor{gray}{0} & \textcolor{gray}{0} & \textcolor{gray}{0} & \textcolor{gray}{0} & \textcolor{gray}{0} & \textcolor{gray}{0} & \textcolor{gray}{0} & \textcolor{gray}{0} & \textcolor{gray}{0} & \textcolor{gray}{0} & \textcolor{gray}{0} & \textcolor{gray}{0} & \textcolor{gray}{0}\\
    O3 & \textcolor{gray}{0} & \textcolor{gray}{0} & +1 & +3 & \textcolor{gray}{0} & \textcolor{gray}{0} & +1 & +2 & \textcolor{gray}{0} & \textcolor{gray}{0} & \textcolor{gray}{0} & \textcolor{gray}{0} & +1 & \textcolor{gray}{0} & \textcolor{gray}{0} & \textcolor{gray}{0} & \textcolor{gray}{0} & \textcolor{gray}{0} & +1 & \textcolor{gray}{0} & \textcolor{gray}{0} & \textcolor{gray}{0} & \textcolor{gray}{0} & +1\\
    Os & \textcolor{gray}{0} & \textcolor{gray}{0} & \textcolor{gray}{0} & \textcolor{gray}{0} & \textcolor{gray}{0} & \textcolor{gray}{0} & \textcolor{gray}{0} & \textcolor{gray}{0} & \textcolor{gray}{0} & \textcolor{gray}{0} & \textcolor{gray}{0} & \textcolor{gray}{0} & \textcolor{gray}{0} & +1 & \textcolor{gray}{0} & \textcolor{gray}{0} & \textcolor{gray}{0} & \textcolor{gray}{0} & +1 & \textcolor{gray}{0} & \textcolor{gray}{0} & \textcolor{gray}{0} & \textcolor{gray}{0} & +1\\
    \midrule
    \multicolumn{25}{c}{Analysis II: AES-DFA} \\ \cmidrule(r){1-1} \cmidrule(lr){2-5} \cmidrule(lr){6-9} \cmidrule(lr){10-15} \cmidrule(lr){16-20} \cmidrule(l){21-25}
    O1 & \textcolor{gray}{0} & \textcolor{gray}{0} & \textcolor{gray}{0} & \textcolor{gray}{0} & \textcolor{gray}{0} & \textcolor{gray}{0} & \textcolor{gray}{0} & \textcolor{gray}{0} & \textcolor{gray}{0} & \textcolor{gray}{0} & \textcolor{gray}{0} & \textcolor{gray}{0} & \textcolor{gray}{0} & \textcolor{gray}{0} & \textcolor{gray}{0} & \textcolor{gray}{0} & \textcolor{gray}{0} & \textcolor{gray}{0} & \textcolor{gray}{0} & \textcolor{gray}{0} & \textcolor{gray}{0} & \textcolor{gray}{0} & \textcolor{gray}{0} & \textcolor{gray}{0}\\
    O2 & \textcolor{gray}{0} & \textcolor{gray}{0} & \textcolor{gray}{0} & \textcolor{gray}{0} & \textcolor{gray}{0} & \textcolor{gray}{0} & \textcolor{gray}{0} & +1 & \textcolor{gray}{0} & \textcolor{gray}{0} & \textcolor{gray}{0} & \textcolor{gray}{0} & \textcolor{gray}{0} & \textcolor{gray}{0} & \textcolor{gray}{0} & \textcolor{gray}{0} & \textcolor{gray}{0} & \textcolor{gray}{0} & \textcolor{gray}{0} & \textcolor{gray}{0} & \textcolor{gray}{0} & \textcolor{gray}{0} & \textcolor{gray}{0} & -1\\
    O3 & \textcolor{gray}{0} & \textcolor{gray}{0} & \textcolor{gray}{0} & \textcolor{gray}{0} & \textcolor{gray}{0} & \textcolor{gray}{0} & \textcolor{gray}{0} & -1 & \textcolor{gray}{0} & \textcolor{gray}{0} & \textcolor{gray}{0} & \textcolor{gray}{0} & \textcolor{gray}{0} & \textcolor{gray}{0} & \textcolor{gray}{0} & -1 & \textcolor{gray}{0} & \textcolor{gray}{0} & -1 & \textcolor{gray}{0} & -1 & \textcolor{gray}{0} & \textcolor{gray}{0} & -1\\
    Os & \textcolor{gray}{0} & \textcolor{gray}{0} & \textcolor{gray}{0} & \textcolor{gray}{0} & \textcolor{gray}{0} & \textcolor{gray}{0} & \textcolor{gray}{0} & -6 & \textcolor{gray}{0} & \textcolor{gray}{0} & \textcolor{gray}{0} & \textcolor{gray}{0} & \textcolor{gray}{0} & \textcolor{gray}{0} & \textcolor{gray}{0} & \textcolor{gray}{0} & \textcolor{gray}{0} & +1 & \textcolor{gray}{0} & \textcolor{gray}{0} & \textcolor{gray}{0} & \textcolor{gray}{0} & +1 & \textcolor{gray}{0}\\
    \bottomrule
\end{tabular}
    \end{adjustbox}
    \caption{Results of Analysis II: Change in exploitable faults by swapping parts of the code in their address spaces.}
    \label{fsim::tab::experiment2::overview}
\end{table*}
Since the size and register usage of the analyzed binaries did not notably change, execution times are similar to \hyperref[fsim::sec::results::analysis1]{Analysis I}.
\Table{fsim::tab::experiment2::overview} shows the results in a condensed form.
The impact of moving code around is visible in both case studies, although the changes in the number of exploitable faults are of different severity in each case study.

For the secure boot case study, the impact is quite significant, since for \texttt{O3}, the number of exploitable faults increases from 17 to 27.
This is due to the exploitability condition:
as soon as the firmware is executed, the fault is regarded as exploitable.
By moving the code around, physical addresses change, which affects how faults can modify control flow.
It is instructive to look at a concrete example from [Secure Boot, \texttt{O1}]:
placing the SHA code in a higher address space than the verification code enables a single-bit instruction fault (fault model \model{8}), which affects the relative offset of a branch instruction.
The new offset happens to point to the very end of the verification code, right where firmware execution is started.
Note that this fault was not introduced by actually changing code, but rather by its final location in the binary.
Typically, a software engineer only specifies the general memory region for all code and the final position of code blocks is managed by the linker.
Hence, this positioning of code blocks is outside of the software engineer's direct scope or control.
Crucially, the resulting change in analyzing physical addresses and offsets can result in new vulnerability that would go unnoticed.

For the AES case study, the largest impact occurs for \texttt{Os} where only 8 faults are affected.
This is a rather insignificant difference, with total values ranging in the thousands.
The small impact despite already large numbers originates from the data-dependent exploitability, which naturally excludes faults that result in major control flow changes.
Since this case study is very sensitive to faults on registers and arithmetic instructions, the physical addresses of instructions are less relevant.

\paragraph{General Observations}
The position of code does have an impact on exploitability.
The change in vulnerability was introduced invisible to the software engineer when branch targets could suddenly be diverted to another useful address.
Notably, this impact can neither be analyzed nor can countermeasures incorporate this information at any layer above machine code, since physical addresses and offsets are not available at those stages.

Systems where faults are exploitable if data is corrupted are much less affected through positioning than systems where the attacker targets control flow.

% ################################################################################
% ################################################################################
% ################################################################################

\subsection{Results of Analysis III -- Influence of Countermeasures}
\label{fsim::sec::results::analysis3}

\fsim took between 1 and 6 minutes (secure boot) and between 3 and 9 seconds (AES-DFA) for each configuration on a single machine utilizing 56 cores.
\Table{fsim::tab::experiment1and3::results} shows the results of our analysis as the relative increase in exploitable faults compared to the setting without countermeasures from \hyperref[fsim::sec::results::analysis1]{Analysis I}.
%Absolute values can be found in \Appendix{fsim::apdx::absolute_results_I_III}.
We verified that both countermeasures successfully defend against the intended attacker models.
This is also reflected in the exploitable faults:
with the instruction replacement technique, there were no exploitable instruction skip faults and with control flow checking, most faults that resulted in illegal branches were detected.

\begin{figure*}[!t]
    \centering
    \begin{subfigure}[b]{0.75\textwidth}
        \centering
        \includegraphics[width=\linewidth]{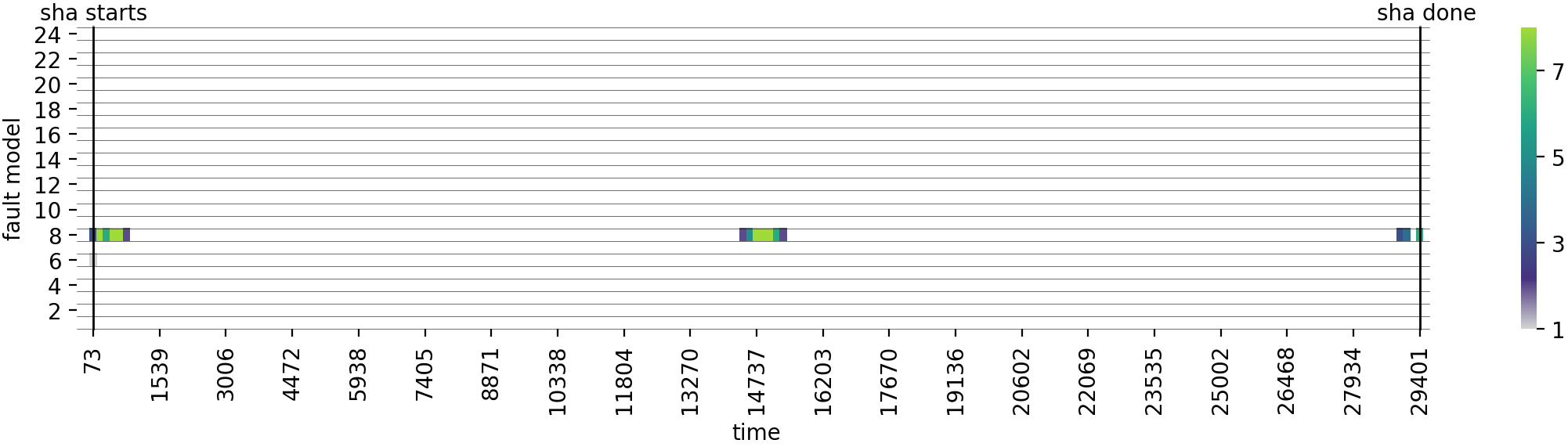}
        \caption{[Secure Boot, \textit{instruction replacement}, \texttt{O2}]}
        \label{fsim::fig::experiment3::sha_vulnerable}
    \end{subfigure}
    \\\ \\
    \begin{subfigure}[b]{0.75\textwidth}
        \centering
        \includegraphics[width=\linewidth]{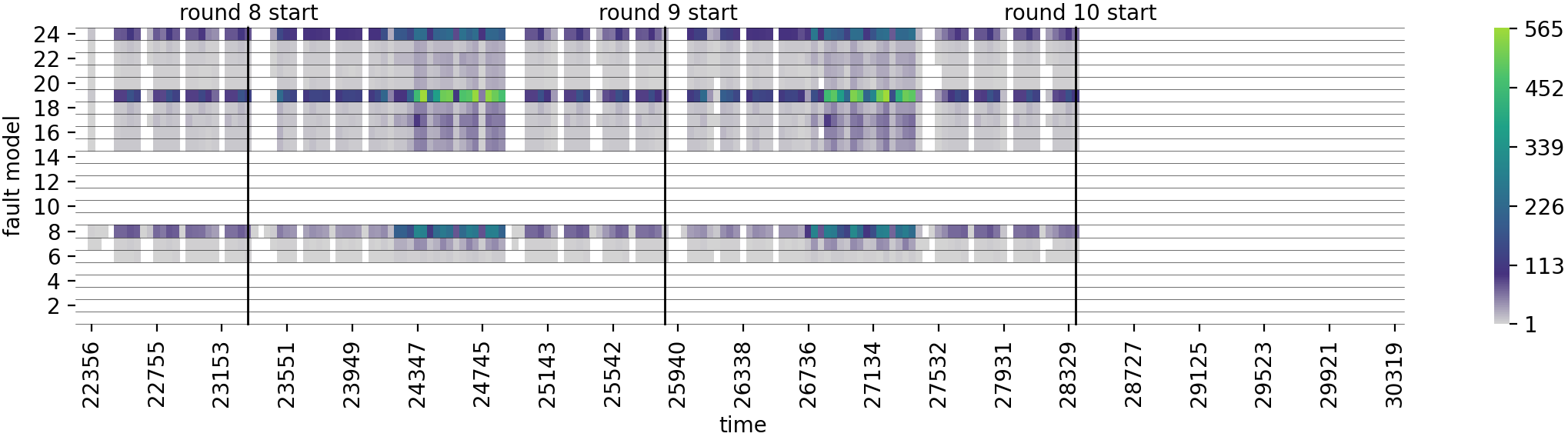}
        \caption{[AES-DFA, \textit{both countermeasures}, \texttt{O1}]}
        \label{fsim::fig::experiment3::aes_vulnerable}
    \end{subfigure}
    \caption{Visualization of non-permanent faults using heatmaps. Whitespace indicates no exploitable faults.}
    \label{fsim::fig::experiment3}
\end{figure*}

In the secure boot case study, the combination of both countermeasures eliminates almost all exploitable faults.
However, applying instruction replacement alone actually \textit{increases} the number of exploitable faults by 428\% - 2812\% compared to the unprotected setting.
More crucially, in the AES-DFA case study, the total number of exploitable faults alarmingly increased by any combination of countermeasures.
To better analyze the issue, we visualized the faults using heatmaps by aggregating the execution time in 200 bins.
Two of these heatmaps are shown in \Figure{fsim::fig::experiment3}.
The heatmap visualizations for all configurations are available \hyperref[fsim::sec::availability]{on GitHub}.

The visualization for [Secure Boot, \textit{instruction replacement}, \texttt{O2}] is shown in \Figure{fsim::fig::experiment3::sha_vulnerable}.
Interestingly, exploitable faults within the SHA computation exist in this and a few other configurations.
Taking a closer look at the numerous faults at the start and middle of the SHA computation, they are mostly similar:
a transient single-bit fault on a branch instruction changes the branch target.
Interestingly, this target now points to the second half of another 32-bit instruction, i.e., its 16 lower bits.
In our case, these 16 bits happen to also be a valid encoding of a 16-bit shift instruction.
Therefore, execution continues and since the faulted branch target is after hash verification, an unauthorized firmware is executed.

For the AES-DFA case study, exploitability increased heavily.
However, there is no direct correlation between code size/execution length and exploitability.
\Figure{fsim::fig::experiment3::aes_vulnerable} shows a heatmap for the \texttt{O1}-optimized code with both countermeasures combined.
For all configurations of the AES-DFA case study, there seems to be a correlation between the success of different fault models:
Whenever numerous single-bit register faults are exploitable (models \model{19} and \model{24}) the number of register faults of the same type that target larger groups of bits also increases.
The same effect is also slightly visible between the number of instruction bit flips (model \model{8}) and related byte-level faults.

\paragraph{General Observations}
Both countermeasures perform perfectly fine within their constrained attacker models.
Regarding secure boot, applying both countermeasures actually results in an almost-secure implementation.
However, in the AES-DFA case study, all countermeasure combinations increased the overall exploitability.
We emphasize that this is not an oversight in the respective countermeasures and that we do not aim to evaluate their strengths or suitability.
Especially since the countermeasures do not aim to protect against data faults, which the AES-DFA case study is highly sensitive to.
Instead, we highlight a major pitfall when applying countermeasures in general by these examples:
through the inclusion of additional instructions, presumably unrelated fault models may substantially gain attack surface.
While the effect of a countermeasure might be verifiable within its target fault model, the effect when faced with other fault models is hard to predict in general and has to be evaluated for each application on a case-by-case basis.

% ################################################################################
% ################################################################################
% ################################################################################

\subsection{Results of Analysis IV -- Injecting Higher-Order Faults}
\label{fsim::sec::results::analysis4}

\begin{figure*}[!t]
    \centering
    \begin{subfigure}[b]{0.48\textwidth}
        \includegraphics[width=\linewidth]{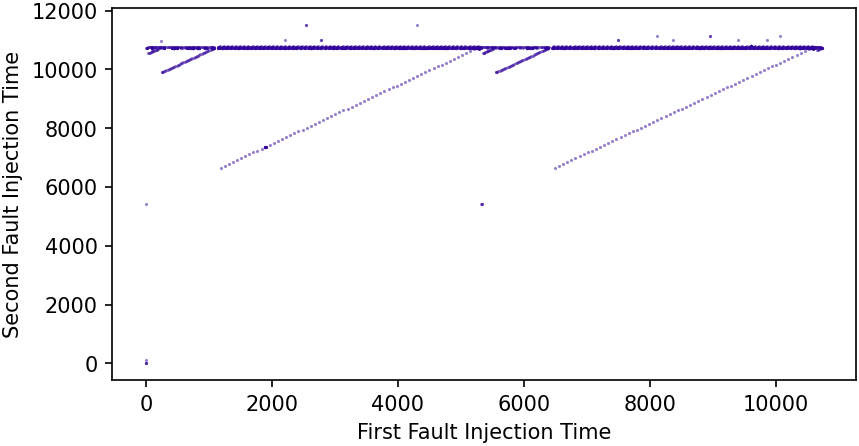}
        \caption{Secure boot case study configuration [\textit{no countermeasures}, \texttt{O2}]}
        \label{fsim::fig::experiment4::bootloader_graph}
    \end{subfigure}
    \hfill
    \begin{subfigure}[b]{0.48\textwidth}
        \includegraphics[width=\linewidth]{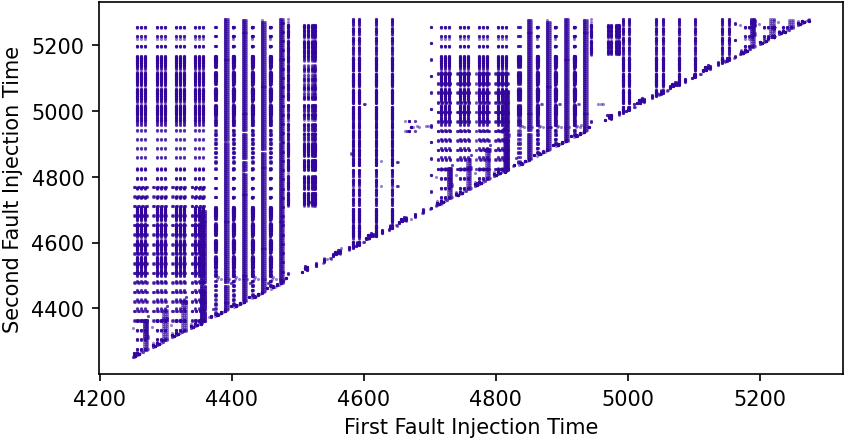}
        \caption{AES-DFA case study configuration [\textit{no countermeasures}, \texttt{O2}]}
        \label{fsim::fig::experiment4::aes_graph}
    \end{subfigure}
    \caption{Visualization of selected results of \hyperref[fsim::sec::results::analysis4]{Analysis IV}. Each dot is an exploitable second-order fault, the $x$-coordinate is the time of the first fault and the $y$-coordinate the time of the second fault. A stronger color indicates more exploitable faults at that location.}
    \label{fsim::fig::experiment4}
\end{figure*}
Exhaustive fault simulation for the \textit{no countermeasures} configurations of the secure boot case study took from 8 hours (\texttt{O3}) up to 24 hours (\texttt{O2}).
For the AES-DFA case study where all configurations were tested, \fsim took from 24 seconds (no countermeasure, \texttt{O3}) up to 3.5 hours (both countermeasures combined, \texttt{O3}).
%A table of the complete results can be found in \Appendix{fsim::apdx::absolute_results_IV}.
\Figure{fsim::fig::experiment4} visualizes the results for two selected configurations, namely (a) [Secure Boot, \textit{no countermeasures}, \texttt{O2}] and (b) [AES-DFA, \textit{no countermeasures}, \texttt{O2}].
Every dot resembles an exploitable second-order fault.
While isolated dots require precise fault injection to exploit, it is interesting to see that both Figures \ref{fsim::fig::experiment4::bootloader_graph} and \ref{fsim::fig::experiment4::aes_graph} contain horizontal or vertical lines.
These lines indicate that, while one fault has to be at a very precise point in time, a second fault within a quite large time range results in exploitable behavior.
This behavior is caused mainly by recurring instructions where the first fault sets a precondition and the second fault causes exploitation (vertical line) or vice versa (horizontal line).
A diagonal line indicates that a fault combination always targets the same instructions/registers but is exploitable at multiple loop iterations, advancing the point in time for both faults each loop.
Interestingly the visualizations directly show where the two SHA blocks are processed or AES rounds are performed.

Looking at the results in detail, we identified multiple unexpected but critical faults.
For example, two transient instruction byte-set faults (model \model{6}) resulted in the following two changes:
First, a \mbox{\texttt{MOV R2, SP}} was faulted to \mbox{\texttt{MOV PC, PC}} which in ARM effectively is a ``skip the \textbf{next} instruction'', since the \texttt{PC} always returns its internal value plus four when read.
This effectively skipped the call to the SHA256 function.
Yet, this fault alone is not exploitable, since the empty, i.e., zero initialized, SHA-output-array would be compared to the array holding the expected hash.
The second fault affects the load of the first byte of the correct hash value, changing \mbox{\texttt{LDRB R1, [R2, \#1]!}} to \mbox{\texttt{LDRB R1, [R2, \#0xff]!}}.
Simplified, this permanently changes the pointer of the correct hash value to an area of RAM that happens to contain only zeroes before the first value is loaded.
Now, the verification code effectively compares two all-zero arrays and thus accepts the unauthorized firmware.
Clearly, such a behavior would be hard to identify, even in a manual inspection of the disassembled machine code.

\paragraph{General Observations}
Exhaustive higher-order fault simulation is a problem limited by its own size.
With an exponentially growing search space the analyst either has to limit the used fault models or invest more computation power and time.
Yet, our results show that an adversary often has a large degree of freedom where to inject his two faults as long as he manages to precisely inject one of them.
This might be especially relevant for jitter- or random-delay-based countermeasures that leverage the assumption that high precision for multiple faults is always required.
We demonstrated that \fsim is capable of uncovering higher-order fault combinations which are highly unlikely to be found by manual inspection, cf.\ the discussed example.
%Furthermore, the vulnerability found in the tests run by us could not be properly assessed without analyzing the physical offsets only available in machine code.

% ################################################################################
% ################################################################################
% ################################################################################

\subsection{Results of Analysis V -- Profiling Undefined Behavior}
\label{fsim::sec::results::analysis5}

In the following, we briefly explain the profiled behavior of the two processors and present the results of using \fsim with a profiling-enhanced \emu.
%Full profiling details are given in \Appendix{fsim::apdx::profiling}.
%Since the same configurations were tested as in \hyperref[fsim::sec::results::analysis3]{Analysis III}, \fsim finished in the same time.

\begin{table*}[!t]
    %\color{blue}
    \centering
    \begin{subfigure}[b]{\textwidth}
        \centering
        \begin{adjustbox}{max width=\textwidth}
            \begin{tabular}{cllc}
                \toprule
                Instruction                           & \multicolumn{1}{c}{Undefined Behavior Condition}              &                                     \multicolumn{2}{c}{Profiled Behavior}                                     \\ \midrule
                \arrayrulecolor{lightgray}
                \texttt{ADD Rdn, Rm} & \texttt{Rdn} == \texttt{Rm} == 15 (PC)                        & hard fault                                                  &      \textcolor{green!40!black}{$\times$}       \\ \midrule
                \texttt{CMP Rn, Rm} (not both from \texttt{R0}-\texttt{R7})   & \texttt{Rn} and \texttt{Rm} both from \texttt{R0}-\texttt{R7} & compares both registers                                     &         \textcolor{orange}{\lightning}          \\ \midrule
                \multirow{2}{*}{\texttt{\{BX/BLX\} Rm}}             & \texttt{Rm} == 15 (PC)                                        & branches to the value of the \texttt{PC} register           &         \textcolor{orange}{\lightning}          \\
                \cmidrule{2-4}                          & last three instruction bits are not 0                         & clears the \texttt{T} bit (thumb mode bit) in \texttt{EPSR} &         \textcolor{orange}{\lightning}          \\ \midrule
                \texttt{POP <register list>}                   & \texttt{register list} empty                                  & equivalent to \texttt{LDR LR, [SP]}, \texttt{SP} unchanged  &      \textcolor{red!80!black}{\lightning}       \\ \midrule
                \texttt{PUSH <register list>}                  & \texttt{register list} empty                                  & equivalent to \texttt{STR LR, [SP]}, \texttt{SP} unchanged  &      \textcolor{red!80!black}{\lightning}       \\ \midrule
                \texttt{LDM Rn! <register list>}                 & \texttt{register list} empty                                  & equivalent to \texttt{LDR LR, [Rn]}, \texttt{Rn} unchanged  &      \textcolor{red!80!black}{\lightning}       \\ \midrule
                \multirow{3}{*}{\texttt{STM Rn! <register list>}}        & \texttt{register list} empty                                  & equivalent to \texttt{STR LR, [Rn], \#4}                    &      \textcolor{red!80!black}{\lightning}       \\
                \cmidrule{2-4}                          & \texttt{Rn} in \texttt{register list} but not lowest          & correctly stores current value of \texttt{Rn}               & \multirow{2}{*}{\textcolor{orange}{\lightning}} \\
                & register in \texttt{register list}                            & and updates \texttt{Rn} in the end                          &                                                 \\
                \arrayrulecolor{black}\bottomrule                &                                                               &                                                             &
            \end{tabular}
        \end{adjustbox}
        \caption{\revise{Profiled undefined behavior of the XMC1100 processor on the Infineon XMC 2Go.}}
        \label{fsim::tab::experiment5::profile_m0}
    \end{subfigure}
    \\\ \\ \ \\
    \begin{subfigure}[b]{\textwidth}
        \centering
        \begin{adjustbox}{max width=\textwidth}
            \begin{tabular}{cllc}
                \toprule
                Instruction                           & \multicolumn{1}{c}{Undefined Behavior Condition}              &                              \multicolumn{2}{c}{Profiled Behavior}                              \\ \midrule
                \arrayrulecolor{lightgray}
                \texttt{ADD Rdn, Rm} & \texttt{Rdn} == \texttt{Rm} == 15 (PC)                        & hard fault                                    &      \textcolor{green!40!black}{$\times$}       \\ \midrule
                \texttt{CMP Rn, Rm} (not both from \texttt{R0}-\texttt{R7})   & \texttt{Rn} and \texttt{Rm} both from \texttt{R0}-\texttt{R7} & correctly compares both registers             &         \textcolor{orange}{\lightning}          \\ \midrule
                \multirow{2}{*}{\texttt{\{BX/BLX\} Rm}}             & \texttt{Rm} == 15 (PC)                                        & hard fault                                    &      \textcolor{green!40!black}{$\times$}       \\
                \cmidrule{2-4}                          & last three instruction bits are not 0                         & correctly executes branch                     &         \textcolor{orange}{\lightning}          \\ \midrule
                \texttt{\{PUSH/POP\} <register list>}              & \texttt{register list} empty                                  & equivalent to \texttt{NOP}                    &         \textcolor{orange}{\lightning}          \\ \midrule
                \texttt{LDM Rn! <register list>}                 & \texttt{register list} empty                                  & equivalent to \texttt{NOP}                    &         \textcolor{orange}{\lightning}          \\ \midrule
                \multirow{3}{*}{\texttt{STM Rn! <register list>}}        & \texttt{register list} empty                                  & equivalent to \texttt{NOP}                    &         \textcolor{orange}{\lightning}          \\
                \cmidrule{2-4}                          & \texttt{Rn} in \texttt{register list} but not lowest          & correctly stores current value of \texttt{Rn} & \multirow{2}{*}{\textcolor{orange}{\lightning}} \\
                & register in \texttt{register list}                            & and updates \texttt{Rn} in the end            &                                                 \\
                \arrayrulecolor{black}\bottomrule                &                                                               &                                               &
            \end{tabular}
        \end{adjustbox}
        \caption{\revise{Profiled undefined behavior of the STM32F407VGT processor.}}
        \label{fsim::tab::experiment5::profile_m4}
    \end{subfigure}
    \caption{\revise{Profiling results of \hyperref[fsim::sec::results::analysis5]{Analysis V}.
        Cases which stop execution do not lead to new faults and, thus, are marked with \textcolor{green!40!black}{$\times$}.
        Those which allow to simply continue emulation in a way that may be anticipated, may surface new faults and are marked with \textcolor{orange}{\lightning}.
        Cases which notably change instruction behavior are marked with \textcolor{red}{\lightning}, since their effects cannot be anticipated by engineers and analysts without profiling.}}
    \label{fsim::tab::experiment5}
    %\color{black}
\end{table*}

\paragraph{XMC1100 Profiling}
Profiling showed that some instructions behave as expected, even in cases of presumably undefined behavior, while others completely change their functionality or simply result in a hard fault.
\revise{
\Table{fsim::tab::experiment5::profile_m0} contains the detailed profiling results for the XMC1000 of all \arm{6} instructions that can potentially result in undefined behavior.
Especially interesting behavior was profiled for the \texttt{PUSH}/\texttt{POP} and \texttt{LDM}/\texttt{STM} instructions if the given register list is empty:
}

\revise{
In that case, \texttt{POP} reads the top of the stack into \texttt{LR} without changing the stack pointer.
Analogously, \texttt{PUSH} overrides the top of the stack with the value in \texttt{LR} without changing the stack pointer.
}

\revise{
\texttt{LDM} reads the value addressed by the base register into the link register \texttt{LR}.
Interestingly though, \texttt{STM} not only overrides the value addressed by the base register with the value in \texttt{LR}, but also increments the value in the base register by 4.
}

The unintuitive behavior of all 4 instructions and their variants may definitely lead to additional exploitable faults, depending on the concrete application.
However, for both case studies (no countermeasures were applied to \arm{6} (thumb1) code), the number of first-order exploitable faults did not change when implementing this behavior in \emu.
This results mostly from the given Assembly code:
the used register lists for PUSH/POP/STM/LDM could most of the time simply not be faulted to be empty with the employed fault models.
The few remaining cases where an empty register list could be provoked, did not lead to exploitable behavior in our code.

\paragraph{STM32F4VGT Profiling}
\revise{
\Table{fsim::tab::experiment5::profile_m4} contains the profiling results of the same instructions that were profiled on the XMC1000.
Interestingly some behavior changed: \texttt{PUSH}, \texttt{POP}, \texttt{LDM}, and \texttt{STM} now simply behave like a \texttt{NOP} in the cases of undefined behavior.
}

Running \fsim with a modified \emu that implements the profiled behavior resulted in the detection of a few additional exploitable faults.
We compared the number of exploitable faults for both case studies and all countermeasure combinations as in \hyperref[fsim::sec::results::analysis3]{Analysis III} for first-order faults.
In the AES-DFA case study, the number of exploitable faults increased by up to 39 faults in 12 of the 16 configurations.
In the secure boot case study only [\textit{instruction replacement}, \texttt{Os}] saw an increase of 6 faults.
However, especially these new faults were very interesting since they did not target any of the profiled instructions directly \revise{but rather were found since \emu did not have to abort because of undefined behavior}.
We used \fsim's fault tracer to inspect said faults in detail:
All 6 faults are instruction bit-flip faults and manipulated an instruction within the SHA to become a branch instruction.
Interestingly, this branch jumps \textbf{behind} the end of our code -- into the literal pool where the constants are stored that are used in SHA.
These constants are now interpreted as instruction encodings and are executed like code.
While the generic \emu quickly reported undefined behavior and exited, the profiled variant was able to continue as it hit one of the profiled edge cases.
Eventually, a part of another SHA constant was decoded as a valid branch, jumping right behind the hash comparison loop and eventually executing the unauthorized firmware.
Again, this is a fault that we are sure would have been missed in manual analysis and could not have been analyzed without physical addresses and a profiled emulator.

\paragraph{General Observations}
While the change in exploitable faults is small in our case studies, this cannot be generalized to other implementations or platforms.
Especially for the STM32F4VGT we only scratched the surface by profiling only a small subset of all potential instructions.
Still, our results show that emulation-based fault simulation definitely misses some faults which can be crucial in real-world applications.
On the other hand, it also demonstrates that \emu can be easily adjusted to incorporate profiled behavior.
%Efficient automated profiling of undefined behavior on real hardware may thus be an interesting direction for future research.
Looking at individual examples, \fsim again uncovered faults that are almost impossible to detect in manual analysis and that are invisible when analyzing only high-level information.

% ################################################################################
% ################################################################################
% ################################################################################

\subsection{Discussion}
We analyzed the impact of various parameters and approaches on exploitability through fault attacks.
Regardless of the specific aspects that were targeted by an individual analysis, a common issue was exploitable faults that stem from the modification of addresses or offsets.
Note that said modifications were induced by all kinds of fault effects, e.g., faults on instructions, operands, immediates, or register values.
We noted several times that the necessary information to analyze such potential vulnerabilities is only available in the final machine code.
Hence, countermeasures applied at any higher code-layer cannot incorporate this data.
Our analyses that involved countermeasures also showed that, while countermeasures may reduce faults in their target domain, the opposite can happen for other fault models -- an increase in vulnerability.

A solution might be to apply countermeasures on the final machine code.
However, modifying the final binary in order to prevent such faults is difficult as well, since inserting instructions at this stage results in a change of numerous address offsets for other instructions, potentially introducing new vulnerabilities, resulting in a vicious cycle.
Therefore, the development of new countermeasures that take machine code information into account is an interesting and important research challenge.

% !TeX root = main.tex

\section{Conclusion}
\label{fsim::sec::conclusion}
In this work, we presented \fsim as a powerful and accessible tool for automated exhaustive fault simulation.
It utilizes \emu as its underlying emulator that we developed specifically with fault injection in mind.
The fault simulator itself is capable to check a large variety of user-defined fault models even at higher orders -- while at the same time keeping the complexity under control through clever shortcuts.

Using \fsim we demonstrated the necessity to incorporate machine code information into analysis and countermeasures.
Fully automated, \fsim detected numerous faults that would have been highly unlikely to be found in manual analysis.
Therefore, we highlight the development of countermeasures against fault injection attacks that take low-level information from machine code into account as an important direction for future research.
With \emu and \fsim available as open-source, analysts and researchers can develop new techniques and automatically verify their effectiveness against a large variety of fault models.
Since our approach is not restricted to security applications but can also be used, e.g., for safety critical code, we regard our results and \fsim as a helpful tool to advance the state of the art.

%-------------------------------------------------------------------------------
\section*{Acknowledgments}
We would like to thank Felix Wegener, Benjamin Kollenda, Phillip Koppe, Marc Fyrbiak, and Bastian Richter for helpful technical discussions.
This work was supported in part by DFG Excellence Strategy grant 39078197 (EXC 2092, CASA) and through ERC grant 695022.

%-------------------------------------------------------------------------------
%\newpage

\bibliographystyle{IEEEtran}
\bibliography{bibliography,localbib}

%-------------------------------------------------------------------------------

\end{document}